\documentclass[twocolumn,twocolappendix]{aastex631}

\newcommand{\grb}{GRB~221009A}
\newcommand{\swift}{\textit{Swift}}
\newcommand{\fermi}{\textit{Fermi}}
\newcommand{\nustar}{\textit{NuSTAR}}

\newcommand{\EK}{\ensuremath{E_{\rm K}}}
\newcommand{\EKiso}{\ensuremath{E_{\rm K,iso}}}

\newcommand{\Egammaiso}{\ensuremath{E_{\gamma,\rm iso}}}	     

\newcommand{\epse}{\ensuremath{\epsilon_{\rm e}}}
\newcommand{\epsb}{\ensuremath{\epsilon_{\rm B}}}
\newcommand{\dens}{\ensuremath{n_{0}}}
\newcommand{\Astar}{\ensuremath{A_{*}}}

\newcommand{\tjet}{\ensuremath{t_{\rm jet}}}
\newcommand{\thetajet}{\ensuremath{\theta_{\rm jet}}}

\newcommand{\pcc}{\ensuremath{{\rm cm}^{-3}}}

\newcommand{\p}{\ensuremath{^{\prime}}}

\newcommand{\nua}{\ensuremath{\nu_{\rm a}}}

\newcommand{\numax}{\ensuremath{\nu_{\rm m}}}
\newcommand{\nuc}{\ensuremath{\nu_{\rm c}}}
\newcommand{\nuaf}{\ensuremath{\nu_{\rm a,f}}}
\newcommand{\numf}{\ensuremath{\nu_{\rm m,f}}}
\newcommand{\nucf}{\ensuremath{\nu_{\rm c,f}}}
\newcommand{\fnumf}{\ensuremath{F_{\nu,\rm m,f}}}
\newcommand{\nuar}{\ensuremath{\nu_{\rm a,r}}}
\newcommand{\numr}{\ensuremath{\nu_{\rm m,r}}}
\newcommand{\nucr}{\ensuremath{\nu_{\rm c,r}}}
\newcommand{\fnumr}{\ensuremath{F_{\nu,\rm m,r}}}
\newcommand{\fnuar}{\ensuremath{F_{\nu,\rm a,r}}}

\newcommand{\nuopt}{\ensuremath{\nu_{\rm opt}}}
\newcommand{\nux}{\ensuremath{\nu_{\rm X}}}

\newcommand{\alphaopt}{\ensuremath{\alpha_{\rm opt}}}
\newcommand{\alphax}{\ensuremath{\alpha_{\rm X}}}
\newcommand{\alphaLAT}{\ensuremath{\alpha_{\rm LAT}}}

\newcommand{\betaniropt}{\ensuremath{\beta_{\rm NIR-opt}}}
\newcommand{\betanirx}{\ensuremath{\beta_{\rm NIR-X}}}

\newcommand{\betax}{\ensuremath{\beta_{\rm X}}}
\newcommand{\betaradio}{\ensuremath{\beta_{\rm radio}}}

\submitjournal{ApJL}
\received{2023 February 04}
\revised{2023 February 15}
\accepted{2023 February 21}

\graphicspath{{./}{figures/}}

\begin{document}

\title{The Radio to GeV Afterglow of GRB 221009A}

\author[0000-0003-1792-2338]{Tanmoy Laskar}
\affiliation{Department of Physics \& Astronomy, University of Utah, Salt Lake City, UT 84112, USA}
\affiliation{Department of Astrophysics/IMAPP, Radboud University, P.O. Box 9010, 6500 GL, Nijmegen, The Netherlands}

\author[0000-0002-8297-2473]{Kate D. Alexander}
\affiliation{Department of Astronomy/Steward Observatory, 933 North Cherry Avenue, Rm. N204, 
Tucson, AZ 85721-0065, USA}

\author[0000-0003-4768-7586]{Raffaella Margutti}
\affil{Department of Astronomy and Physics, University of California, 501 Campbell Hall, Berkeley, CA 94720-3411, USA}

\author[0000-0003-0307-9984]{Tarraneh Eftekhari}\thanks{NHFP Einstein Fellow}
\affiliation{Department of Physics and Astronomy, Northwestern University, 2145 Sheridan Road, Evanston, IL 60208-3112, USA}
\affiliation{Center for Interdisciplinary Exploration and Research in Astrophysics, 1800 Sherman Avenue, Evanston, IL 60201, USA}

\author[0000-0002-7706-5668]{Ryan Chornock}
\affil{Department of Astronomy, University of California, 501 Campbell Hall, Berkeley, CA 94720-3411, USA}


\author[0000-0002-9392-9681]{Edo Berger}
\affiliation{Center for Astrophysics | Harvard \& Smithsonian, Cambridge, MA 02138, USA}

\author[0000-0001-7007-6295]{Yvette Cendes}
\affiliation{Center for Astrophysics | Harvard \& Smithsonian, Cambridge, MA 02138, USA}

\author[0000-0003-1716-4119]{Anne Duerr}
\affiliation{Department of Physics \& Astronomy, University of Utah, Salt Lake City, UT 84112, USA}

\author[0000-0001-8472-1996]{Daniel A. Perley}
\affiliation{Astrophysics Research Institute, Liverpool John Moores University, IC2, Liverpool Science Park, 146 Brownlow Hill, Liverpool L3 5RF, UK}

\author[0000-0003-3193-4714]{Maria Edvige Ravasio}
\affiliation{Department of Astrophysics/IMAPP, Radboud University, P.O. Box 9010, 6500 GL, Nijmegen, The Netherlands}
\affiliation{INAF--Astronomical Observatory of Brera, via E. Bianchi 46, I-23807 Merate, Italy}

\author[0000-0002-1251-7889]{Ryo Yamazaki}
\affiliation{Department of Physical Sciences, Aoyama Gakuin University, 5-10-1 Fuchinobe, Sagamihara 252-5258, Japan}
\affiliation{Institute of Laser Engineering, Osaka University, 2-6 Yamadaoka, Suita, Osaka 565-0871, Japan}


\author[0000-0002-8225-5431]{Eliot H. Ayache} 
\affiliation{The Oskar Klein Centre, Department of Astronomy, Stockholm University, AlbaNova, SE-106 91 Stockholm, Sweden}

\author[0000-0001-7139-2724]{Thomas Barclay}
\affiliation{NASA Goddard Space Flight Center, Greenbelt, MD 20771, USA}
\affiliation{University of Maryland, Baltimore County, 1000 Hilltop Cir, Baltimore, MD 21250, USA}

\author[0000-0002-5565-4824]{Rodolfo Barniol Duran}
\affiliation{Department of Physics and Astronomy, California State University, Sacramento, 6000 J Street, Sacramento CA 95819-6041 USA}

\author[0000-0003-3460-506X]{Shivani Bhandari}
\affiliation{ASTRON, Netherlands Institute for Radio Astronomy, Oude Hoogeveensedijk 4, NL-7991 PD Dwingeloo, the Netherlands}
\affiliation{Joint institute for VLBI ERIC, Oude Hoogeveensedijk 4, NL-7991 PD Dwingeloo, the Netherlands}
\affiliation{Anton Pannekoek Institute for Astronomy, University of Amsterdam, Science Park 904, NL-1098 XH Amsterdam, the Netherlands}
\affiliation{CSIRO, Space and Astronomy, P.O. Box 76, Epping, NSW 1710 Australia}

\author[0000-0001-6415-0903]{Daniel Brethauer}
\affiliation{Department of Astronomy, University of California, 501 Campbell Hall, Berkeley, CA 94720-3411, USA}

\author[0000-0003-0528-202X]{Collin T. Christy}
\affiliation{Department of Astronomy/Steward Observatory, 933 North Cherry Avenue, Rm. N204, 
Tucson, AZ 85721-0065, USA}

\author[0000-0001-5126-6237]{Deanne L. Coppejans}
\affiliation{Department of Physics, University of Warwick, Coventry CV4 7AL, UK}

\author[0000-0001-7626-9629]{Paul Duffell}
\affiliation{Department of Physics and Astronomy, Purdue University, 525 Northwestern Avenue, West Lafayette, IN 47907, USA}

\author[0000-0002-7374-935X]{Wen-fai Fong}
\affiliation{Department of Physics and Astronomy, Northwestern University, 2145 Sheridan Road, Evanston, IL 60208-3112, USA}
\affiliation{Center for Interdisciplinary Exploration and Research in Astrophysics, 1800 Sherman Avenue, Evanston, IL 60201, USA}

\author[0000-0002-0908-914X]{Andreja Gomboc}
\affiliation{Center for Astrophysics and Cosmology, University of Nova Gorica, Vipavska 11c, 5270 Ajdov\v s\v cina, Slovenia}

\author[0000-0001-6869-0835]{Cristiano Guidorzi}
\affiliation{Department of Physics and Earth Science, 
University of Ferrara, Via Saragat 1, I-44122 Ferrara, Italy}
\affiliation{INFN -- Sezione di Ferrara, Via Saragat 1, 44122 Ferrara, Italy}
\affiliation{INAF -- Osservatorio di Astrofisica e Scienza dello Spazio di Bologna, Via Piero Gobetti 101, 40129 Bologna, Italy}

\author[0000-0002-6745-4790]{Jamie A. Kennea}
\affiliation{Department of Astronomy and Astrophysics, The Pennsylvania State University, 525 Davey Lab, University Park, PA 16802, USA}

\author[0000-0001-7946-4200]{Shiho Kobayashi}
\affiliation{Astrophysics Research Institute, Liverpool John Moores University, IC2, Liverpool Science Park, 146 Brownlow Hill, Liverpool L3 5RF, UK}

\author[0000-0001-7821-9369]{Andrew Levan}
\affiliation{Department of Astrophysics/IMAPP, Radboud University, P.O. Box 9010, 6500 GL, Nijmegen, The Netherlands}

\author[0000-0003-1622-1484]{Andrei P.\ Lobanov}
\affiliation{Max-Planck-Institut f\"ur Radioastronomie, Auf dem H\"ugel 69, D-53121 Bonn, Germany}

\author[0000-0002-4670-7509]{Brian D. Metzger}
\affiliation{Department of Physics and Columbia Astrophysics Laboratory, Columbia University, Pupin Hall, New York, NY 10027, USA}
\affiliation{Center for Computational Astrophysics, Flatiron Institute, 162 5th Ave, New York, NY 10010, USA} 
\author[0000-0001-9503-4892]{Eduardo Ros}
\affiliation{Max-Planck-Institut f\"ur Radioastronomie, Auf dem H\"ugel 69, D-53121 Bonn, Germany}

\author[0000-0001-9915-8147]{Genevieve~Schroeder}
\affiliation{Department of Physics and Astronomy, Northwestern University, 2145 Sheridan Road, Evanston, IL 60208-3112, USA}
\affiliation{Center for Interdisciplinary Exploration and Research in Astrophysics, 1800 Sherman Avenue, Evanston, IL 60201, USA}

\author[0000-0003-3734-3587]{P.\ K.\ G.\ Williams}
\affiliation{Center for Astrophysics | Harvard \& Smithsonian, Cambridge, MA 02138, USA}

\shorttitle{The Radio to GeV Afterglow of GRB 221009A}
\shortauthors{Laskar et al.}

\begin{abstract}

\grb\ ($z=0.151$) is one of the closest known long $\gamma$-ray bursts (GRBs). Its extreme brightness across all electromagnetic wavelengths provides an unprecedented opportunity to study a member of this still-mysterious class of transients in exquisite detail. We present multi-wavelength observations of this extraordinary event, spanning 15 orders of magnitude in photon energy from radio to $\gamma$-rays. We find that the data can be partially explained by a forward shock (FS) from a highly-collimated relativistic jet interacting with a low-density wind-like medium. {Under this model, the} jet's beaming-corrected kinetic energy ($E_K \sim 4\times10^{50}$ erg) is typical for the GRB population. The radio and mm data provide strong limiting constraints on the FS model, but require the presence of an additional emission component. From equipartition arguments, we find that the radio emission is likely produced by a small amount of mass ($\lesssim6\times10^{-7} M_\odot$) moving relativistically ($\Gamma\gtrsim9$) with a large kinetic energy ($\gtrsim10^{49}$~erg). However, the temporal evolution of this component does not follow prescriptions for synchrotron radiation from a single power-law distribution of electrons (e.g.\ in a reverse shock or two-component jet), or a thermal electron population, perhaps suggesting that one of the standard assumptions of afterglow theory is violated. \grb\ will likely remain detectable with radio telescopes for years to come, providing a valuable opportunity to track the full lifecycle of a powerful relativistic jet.
\end{abstract}

\section{Introduction}
\label{text:intro}
Long-duration $\gamma$-ray bursts (GRBs) release enormous amounts of energy in the form of collimated, highly-relativistic jets. These outflows are thought to be launched by a powerful magnetar or accreting black hole born during the collapse of a massive star \citep{pir05,wb06,mab+11}. GRBs are typically discovered via their prompt (seconds -- minutes duration) $\gamma$-ray emission, possibly powered by internal shocks within the jet \citep{rm92,kps97,kp00a}. Properties of the jet, such as its structure, energetics, and magnetization, along with other physically interesting quantities such as the structure and density of the ambient medium, can be derived from modeling the broadband afterglow emission, which typically remains detectable in the radio, optical, and X-ray bands for days to months (e.g., \citealt{spn98}).

Synchrotron models, based on both analytical and numerical or hydrodynamic calculations, have been remarkably successful in explaining the multi-wavelength afterglow emission of GRBs. In the basic picture, the afterglow is modeled as synchrotron radiation produced by a population of relativistic electrons accelerated in the forward shock (FS) produced by the interaction of the jet with the ambient medium. The radiating electrons are assumed to be accelerated into a single power-law distribution of energies, characterized by a minimum energy, $\gamma_m m_e c^2$ and a power-law index $p$. This creates a simple broken power-law emission spectrum, fully characterized by a few break frequencies and overall flux normalization. 
Physical model parameters such as the jet energy and opening angle, the microphysical shock parameters, and the density and density profile of the circumburst medium 
may be determined by capturing the full synchrotron SED and its temporal evolution, which typically requires observations spanning the full electromagnetic spectrum.

Despite the success of this basic picture, increasingly detailed observational campaigns of GRBs over the past decade have demonstrated the need to incorporate additional physics. One of the most important additions has been the detection of reverse shock (RS) emission, a second synchrotron component from a second shockwave propagating back into the jet \citep{lbz+13,pcc+14,vdhpdb+14}. The RS emission reveals details of the jet's magnetization and initial bulk Lorentz factor \citep{sp99, kz03, zk05}. Additional model extensions, from the inclusion of multiple ejecta shells moving at different Lorentz factors \citep{rm98,jbg06,lbm+15} to the consideration of emission from electrons not accelerated in the FS (thermal electrons; \citealt{ew05,rl17,wbi+18,mq21}), have also been explored in the literature. Further extensions may be needed to elucidate several aspects of GRB observations that remain unexplained. These include the detection of very high-energy photons ($>1$ TeV) associated with some GRBs \citep{aaa+19,mcaa+19a,gcn29075,hcaa+21} and unusual radio evolution inconsistent with variability due to extrinsic scattering from the interstellar medium or the standard synchrotron emission framework \citep{fmb+04,bhvdh+19,kf21,ldf+23}. Numerical simulations may provide the ultimate solution to these oddities, but rely on uncertain assumptions about particle acceleration physics, magnetic field structure, etc.

On 2022 October 9, the Fermi Gamma-Ray Burst Monitor (GBM) and the Swift Burst Alert Telescope (BAT) triggered on a new $\gamma$-ray source, rapidly confirmed to be the brightest GRB ever seen by these instruments and designated \grb\ \citep{gcn32632,gcn32636,wkd+23}. These detections were followed by numerous others across the electromagnetic spectrum, confirming a bright optical, X-ray, and radio counterpart \citep{gcn32632,gcn32634,gcn32635,gcn32653} and a redshift of 0.151 (\citep{gcn32648,mli+23}). \grb\ also exhibited coincident very high-energy emission, with photons reported up to 18 TeV \citep{gcn32677} and potentially as high as $250$~TeV \citep{ATel15669}. Although a calculation of the isotropic-equivalent energy released in $\gamma$-rays for this burst is complicated by saturation effects at multiple $\gamma$-ray instruments, preliminary estimates yield $\log\Egammaiso\approx54.77$ (0.1~keV--100~MeV), one of the highest values to date \citep{gcn32762}.

Several groups have modeled the afterglow emission from \grb\ \citep{rwz22, smoy22, gtbr23, fsr+23}. \cite{fsr+23} focus solely on the optical and X-ray afterglow in the context of characterizing a possible emerging supernova (SN) component at $\sim20$ days. The other three groups use radio, optical, X-ray, and/or $\gamma$-ray observations taken at $\lesssim12$ days post-discovery\footnote{{Based on information available at the time of submission of this manuscript.}} \citep{rwz22, smoy22, gtbr23}. Although the details of each model differ, common findings among the three groups include a high isotropic-equivalent kinetic energy and a high degree of collimation for the jet and a low-density environment. In all analyses, the best-fit model cannot explain all of the data, suggesting that \grb\ may be an ideal test case for more complex models incorporating more realistic physics.

Here, we present detailed radio, millimeter, optical, X-ray, and $\gamma$-ray observations of \grb, spanning fifteen orders of magnitude in frequency and four orders of magnitude in time (extending to $\sim100$ d). We find that a standard synchrotron FS model can explain large portions of the data, revealing that \grb's extraordinary brightness is due to the jet's unusual degree of collimation, rather than an intrinsically large energy. However, the model struggles to reproduce some aspects of the full dataset, in particular the radio and millimeter emission. We show that standard extensions to the synchrotron FS model (e.g., the addition of a reverse shock) are unsuccessful at explaining our observations, suggesting that at least one of the basic assumptions underlying our standard picture of GRB afterglow emission needs to be revisited. 

Unless otherwise noted, all times refer to times after the GBM trigger (2022 Oct 09 13:16:59.000 UTC; \citealt{gcn32636}), and all magnitudes are in the AB system, not corrected for Galactic or intrinsic dust extinction. We employ $\Lambda$CDM cosmological parameters of $\Omega_m = 0.31$, $\Omega_\Lambda = 0.69$, and $H_0 = 68$~km\,s$^{-1}$\,Mpc$^{-1}$ throughout, and assume a Galactic extinction of $A_V=4.1034$~mag \citep{sf11}. The redshift of $z=0.151$ corresponds to a luminosity distance of $D_{\rm L}=2.28\times10^{27}$~cm for this burst. 

\section{Observations and Data Analysis}
\subsection{Radio}

\subsubsection{GMRT}
\noindent We observed the afterglow using the upgraded Giant Metrewave Radio Telescope (GMRT) through program 43\_039 (PI: Laskar) starting 18.0~days, 21.9~days, and 21.8~days after the burst in Bands 3 ($400$~MHz), 4 (750~MHz), and 5 (1260~MHz), respectively. The observations utilized 400~MHz bandwidth. We used J1925+2106 for complex gain calibration, and either 3C~286 or 3C~48 for bandpass and flux density calibration, depending on availability. For Band 3, we carried out data reduction via standard techniques using the Common Astronomy Software Applications (CASA; \citealt{mws+07}), including three rounds of phase self-calibration, followed by two rounds of amplitude and phase self-calibration. For Band 4 and Band 5, we reduced and imaged the data using the \texttt{capture} pipeline \citep{ki21}. We list the results of our uGMRT observations in Table~\ref{tab:data:radio}. 

\subsubsection{MeerKAT}
\noindent We observed the afterglow with the MeerKAT radio telescope beginning at $\approx1.3$~days after the burst in L Band (1.3~GHz) via program  SCI-20220822-TL-01 (PI: Laskar). Later observations employed simultaneous subarray-mode observations in the L and UHF bands. The observations employed 1938-634 as bandpass and flux density calibrator, and J1925+2106 as complex gain calibrator. The pipeline SDP images were of sufficient quality for photometry of the afterglow using CASA. The noise in the image of L-band data taken on 2022 Oct 15 was much higher than in the other epochs due to significant RFI contamination. 
We report our MeerKAT photometry in Table~\ref{tab:data:radio}. 

\subsubsection{VLA}
\noindent We obtained seven epochs of multi-frequency observations of \grb\ beginning $\approx3.5$~days after the burst with the Karl G.~Jansky Very Large Array (VLA) through program 22B\_062 (PI: Laskar). The observations employed either 3C~286 or 3C~48 for bandpass and flux density calibration and J1925+2106 for complex gain calibration. We imaged the pipeline-calibrated measurement sets downloaded from the VLA archive, where available. For the second epoch taken on 2022 Oct 15, the flux calibrator scans were defective. We calibrated this observation using flux calibrator data at the same frequencies taken from the first epoch instead, and achieved excellent agreement for the derived flux density of the gain calibrator. The pattern of the imaging residuals in the second and third epochs indicated gain errors, which we ameliorated via phase and amplitude self-calibration using an automated self-calibration pipeline\footnote{\url{https://github.com/jjtobin/auto_selfcal}}. We report the results of our VLA observations in Table~\ref{tab:data:radio}. 

\subsubsection{VLBA}
 We obtained four epochs of observations of \grb\ with the Very Long Baseline Array (VLBA) and the 100-m radio telescope Effelsberg at 8.3~GHz beginning $\approx12.4$~days after the burst under project VLBA/22B-305 (Legacy Code \texttt{BL073}; PI: Laskar). The full-track observations utilized 3C~345 and 3C~454.3 as fringe calibrators and were phase referenced to TXS\,1903+196 (JVAS\,J1905+1943), a compact quasar at $z=2.3$, with cycles of 168\,s on target and 68\,s on the calibrator. 
We carried out standard fringe fitting\footnote{To account for the structure of the complex gain calibrator, we derived a clean-component model by first imaging the data after a preliminary fringe fit and used this model for subsequent calibrations.}, bandpass, and complex gain calibration using \texttt{AIPS} \citep{gre03}. We interpolated the delay and rate solutions to \grb\ and produced phase-reference images. We measured the flux density using \textsc{jmfit} in \texttt{AIPS} in the image plane and with \textsc{modelfit} in \textsc{Difmap} \citep{spt94} in the $uv$ plane, obtaining similar results within a few percent. We corrected the derived flux densities for the primary beam for the first epoch and re-scaled according to the calibrator flux in the fourth epoch. We present the inferred total (CLEANed) flux density measurements from our VLBA observations in Table~\ref{tab:data:radio}. 

\subsubsection{ATCA}
\noindent We obtained four epochs of observations of the afterglow with the Australia Telescope Compact Array (ATCA) beginning $\approx5.8$~days after the burst by triggering our program C3289 (PI: Laskar) using the CABB correlator tuned to 15~mm (with 2~GHz basebands tuned to 16.7 and 21.2~GHz) and 7~mm (33 and 35 GHz; and also 45 and 47~GHz). The observations were carried out by the observatory in service mode under project CX515, and employed PKS1921-293 for bandpass calibration, PKS1934-638 for flux density calibration, and PKS1923+210 (J1925+2106) for complex gain calibration. We calibrated the data using \texttt{MIRIAD} \citep{stw95}, and combined the 33~GHz and 35~GHz data, as well as the 45~GHz and 47~GHz data, prior to imaging in CASA. The cleaning process reveals residuals characteristic of phase decorrelation and the resulting ATCA SEDs are extremely steep and are inconsistent with an extrapolation to nearly contemporaneous ALMA observations. Whereas phase-only self-calibration recovers some flux, the signal-to-noise in the data is too low for adequate 
self-calibration. We list the ATCA measurements in Table~\ref{tab:data:radio} for completeness, but we do not use these in our subsequent modeling, and we caution against the use of these data in other works without more careful attention to the calibration.

\subsubsection{ALMA}
We obtained seven epochs of ALMA Band 3 (3~mm) observations of \grb\ beginning $\approx2.4$ days after the burst through program 2022.1.01433.T (PI: Laskar). The observations utilized two 4 GHz wide basebands centered at 91.5 and 103.5 GHz, respectively. The first five epochs used J1924-2914 as flux density calibrator. The sixth epoch used J1550+0527; this execution was affected by a correlator issue and was re-observed, this time using J2232+1143. The seventh epoch used J1550+0527. All epochs used J1914+1636 as complex gain calibrator. Our ALMA coverage has a gap between $\approx15$--75~days due to the shutdown of the observatory following a cyber attack. We downloaded the pipeline-generated images from the ALMA archive and performed photometry in CASA. We report our photometry in Table~\ref{tab:data:radio}. 

\subsubsection{NOEMA}
We obtained two epochs of NOEMA 97.5\,GHz observations of \grb\ at 39.2 and 54.2 days after the burst through program S22BE (PI: Laskar). The observations utilized two 7.7~GHz wide basebands centered at 89.8~GHz and 105.2~GHz, respectively. We used MWC349 as flux density calibrator, B2200+420 (epoch 1) and 3C~454.3 (epoch 2) as bandpass calibrators, and 1932+204 and 1923+210 as complex gain calibrators. Data were reduced in GILDAS\footnote{\url{https://www.iram.fr/IRAMFR/GILDAS}} using standard procedures by observatory staff and provided to us. We imaged the reduced data in CASA and report our photometry in Table~\ref{tab:data:radio}. 

\subsubsection{SMA}
We observed the afterglow with the Submillimeter Array (SMA) at a combination of 1.3~mm ($\approx230$~GHz) and 1.1~mm ($290$~GHz) for 7 epochs using Uranus as flux density calibrator, 3C~84 as bandpass calibrator, and interleaved observations of J1925+211 and MWC349a for complex gain calibration. We calibrated the data in MIR\footnote{\url{https://lweb.cfa.harvard.edu/rtdc/SMAdata/process/mir}} (the in-house calibration suite for the SMA) and measured the afterglow flux density using vector averaging of the $uv$ data (verified by imaging of the first two epochs). An additional 850~$\mu$m observation yielded an upper limit. We report the results of our SMA observations in Table~\ref{tab:data:radio}.

\begin{deluxetable}{ccccc}
 \tabletypesize{\footnotesize}
 \tablecolumns{5}
 \tablecaption{Radio observations of \grb}
 \tablehead{   
   \colhead{Telescope} &
   \colhead{Frequency$^{\rm{a}}$} &
   \colhead{Time$^{\rm{b}}$} &
   \colhead{Flux density} &
   \colhead{Uncertainty}\\
   \colhead{} &
   \colhead{(GHz)} &
   \colhead{(days)} &
   \colhead{(mJy)} &
   \colhead{($\mu$Jy)}
   }
 \startdata 
MeerKAT & 1.28 & 1.26 & 2.10 & 24 \\
ALMA    & 97.5 & 2.36 & 9.21 & 90 \\
SMA     & 231  & 2.68 & 9.44 & 900 \\
\ldots & \ldots & \ldots & \ldots & \ldots
\enddata
\tablecomments{$^{\rm{a}}$ Central frequency. $^{\rm{b}}$ Mid-time since \fermi/GBM trigger. 
The full data table is available {as an enhanced machine-readable table on-line.}}
\label{tab:data:radio}
\end{deluxetable}

\subsection{Optical}
We observed \grb\ in $griz$ filters with IO:O on the Liverpool Telescope (LT; \citealt{ssr+04}) at multiple epochs beginning on 2022 Oct 09. We downloaded pipeline-reduced images from the LT archive. To avoid contamination from nearby sources in the crowded field, we use a custom script to construct a model of the PSF and subtract the wings of several bright stars after masking the inner pixels, as needed. We report aperture photometry performed in a $1.0\arcsec$ radius aperture calibrated to the Pan-STARRS1 catalog in Table~\ref{tab:data:lt}. 
We additionally include optical and near-infrared (NIR) data reported in GCN circulars, from 
\cite{gcn32647,gcn32652,gcn32749,gcn32750,gcn32755,gcn32758,gcn32804,gcn32860}
and radio observations from \cite{gcn32676} and \cite{gcn32736} 
in our analysis. We do not include \textit{Hubble Space Telescope} (\textit{HST}) observations from \cite{gcn32921} in our work as the corresponding photometry likely contains contamination from an underlying host galaxy, and require deep, late-time templates for subtraction. 

\subsection{Ultraviolet}
We performed photometry in a 5\arcsec\, aperture on all \swift/UVOT images up to and including segment 01126853067 (taken on 2022 Dec 16) obtained from the \swift\ repository\footnote{http://www.swift.ac.uk/swift\_portal} with the \texttt{uvotproduct} (v2.8) software and CALDB version 20221229. This corresponds to almost all of the data taken before the target entered a Sun constraint on 2022 Dec 21 (lasting until 2023 Feb 06). We used defaults for all pipeline parameters. We present our UVOT measurements in Table~\ref{tab:data:uvot} (see also \citealt{wkd+23} for an independent analysis of these data). The \textit{white}-band data is of limited utility owing to the strong foreground extinction and we therefore do not use data in this band for subsequent modeling.

\startlongtable
\begin{deluxetable}{ccccccc}
\tabletypesize{\scriptsize}
\tablecaption{Liverpool Telescope Observations of \grb\ \label{tab:data:lt}}
\tablehead{
\colhead{$\Delta t$ $^{\rm{a}}$} & \colhead{Filter} & \colhead{Mag} & \colhead{Uncertainty} & \colhead{$t_{\rm int}$} & Seeing \\ 
\colhead{(d)} & \colhead{} & \colhead{(AB)} & \colhead{} & \colhead{(s)} & \colhead{(arcsec)} }
\startdata
 0.33486 & g & 18.57 &  0.04 &   45 & 1.17 \\
 0.33749 & z & 15.29 &  0.01 &   45 & 1.04 \\
 0.33662 & i & 15.96 &  0.01 &   45 & 1.08 \\
 0.33576 & r & 16.99 &  0.01 &   45 & 1.40 \\
 0.34899 & g & 18.53 &  0.04 &   50 & 1.01 \\
 0.34987 & r & 17.05 &  0.01 &   40 & 0.99 \\
 0.35061 & i & 16.00 &  0.01 &   30 & 0.90 \\
 0.35132 & z & 15.33 &  0.01 &   30 & 0.89 \\
 0.39745 & g & 18.55 &  0.05 &   50 & 1.36 \\
 0.39978 & z & 15.48 &  0.01 &   30 & 1.02 \\
 0.39834 & r & 17.21 &  0.01 &   40 & 1.01 \\
 0.39908 & i & 16.15 &  0.01 &   30 & 0.96 \\
 1.27230 & r & 18.85 &  0.02 &   45 & 0.87 \\
 1.27097 & g & 20.33 &  0.06 &  100 & 1.34 \\
 1.27404 & z & 17.06 &  0.01 &   45 & 1.03 \\
 1.27316 & i & 17.77 &  0.01 &   45 & 0.84 \\
 1.30908 & r & 18.87 &  0.02 &   90 & 1.08 \\
 1.31046 & i & 17.80 &  0.01 &   90 & 1.03 \\
 1.31186 & z & 17.10 &  0.01 &   90 & 1.00 \\
 1.39474 & g & 20.41 &  0.12 &  100 & 1.17 \\
 1.39607 & r & 19.03 &  0.04 &   45 & 1.06 \\
 1.39694 & i & 17.93 &  0.01 &   45 & 1.41 \\
 1.39781 & z & 17.21 &  0.01 &   45 & 1.14 \\
 2.26839 & i & 18.69 &  0.02 &  100 & 1.11 \\
 2.26689 & r & 19.84 &  0.05 &  100 & 1.35 \\
 2.26991 & z & 18.02 &  0.01 &  100 & 1.02 \\
 2.29593 & r & 19.84 &  0.05 &   60 & 1.04 \\
 2.29441 & g & 21.05 &  0.12 &  120 & 1.33 \\
 2.29698 & i & 18.73 &  0.02 &   60 & 1.24 \\
 2.29803 & z & 17.99 &  0.02 &   60 & 1.24 \\
 3.27665 & i & 19.29 &  0.03 &  120 & 1.51 \\
 3.27434 & r & 20.31 &  0.05 &  180 & 1.58 \\
 3.27863 & z & 18.43 &  0.07 &  120 & 2.69 \\
 3.28182 & g & 21.61 &  0.19 &  180 & 2.18 \\
 3.28371 & r & 20.33 &  0.09 &   60 & 1.91 \\
 3.28475 & i & 19.28 &  0.03 &   60 & 1.46 \\
 3.28579 & z & 18.55 &  0.03 &   60 & 1.56 \\
 7.27887 & r & 21.60 &  0.08 &  180 & 0.79 \\
 7.28154 & i & 20.44 &  0.03 &  180 & 0.85 \\
 7.28571 & z & 19.68 &  0.03 &  360 & 0.75 \\
 7.28992 & i & 20.37 &  0.03 &  180 & 0.83 \\
 7.29263 & r & 21.44 &  0.08 &  180 & 0.82 \\
 8.27806 & i & 20.58 &  0.04 &  240 & 1.07 \\
 8.27470 & r & 21.68 &  0.08 &  240 & 1.07 \\
 8.28328 & z & 19.94 &  0.04 &  480 & 1.20 \\
 8.28854 & i & 20.69 &  0.05 &  240 & 1.08 \\
 8.29195 & r & 21.91 &  0.13 &  240 & 1.09 \\
10.28995 & r & 22.04 &  0.11 &  300 & 0.90 \\
10.29366 & i & 21.07 &  0.06 &  240 & 0.84 \\
10.29887 & z & 20.27 &  0.05 &  480 & 1.04 \\
10.30412 & i & 20.92 &  0.05 &  240 & 0.85 \\
10.30788 & r & 22.04 &  0.10 &  300 & 0.86 \\
11.27403 & r & 22.09 &  0.11 &  360 & 0.83 \\
11.28344 & z & 20.50 &  0.05 &  480 & 0.80 \\
11.27821 & i & 21.12 &  0.05 &  240 & 0.83 \\
11.28871 & i & 21.17 &  0.06 &  240 & 0.91 \\
11.29292 & r & 22.09 &  0.10 &  360 & 0.91 \\
14.28998 & r & 22.45 &  0.17 &  360 & 0.90 \\
14.29496 & i & 21.47 &  0.07 &  360 & 0.87 \\
14.30262 & z & 20.83 &  0.06 &  720 & 0.79 \\
14.31111 & i & 21.35 &  0.09 &  240 & 0.95 \\
14.31535 & r & 22.35 &  0.15 &  360 & 0.99 \\
17.31003 & z & 21.08 &  0.15 &  720 & 1.67 \\
18.29124 & i & 21.91 &  0.13 &  720 & 1.37 \\
18.30665 & r & 22.99 &  0.24 & 1350 & 1.46 \\
19.27566 & z & 21.19 &  0.73 &  600 & 1.08 \\
20.27930 & z & 21.55 &  0.14 & 1350 & 1.23 \\
21.28883 & z & 21.31 &  0.12 & 1350 & 1.23 \\
36.28571 & i & 22.23 &  0.11 & 1620 & 1.10 \\
38.27709 & r & 24.06 &  0.32 & 1800 & 0.87 \\
\enddata
\tablecomments{$^{\rm{a}}$Mid-time since \fermi/GBM trigger. The data have not been corrected for extinction in the Milky Way or GRB host galaxy, or for the contribution of host galaxy light.}
\end{deluxetable}

\begin{deluxetable}{ccccc}
\tabletypesize{\scriptsize}
\tablecaption{\swift/UVOT Observations of \grb\ \label{tab:data:uvot}}
\tablehead{
\colhead{Start Time$^{\rm{a}}$} & \colhead{Stop Time$^{\rm{a}}$} & \colhead{Band} & \colhead{Mag} & \colhead{Uncertainty}\\ 
\colhead{(s)} & \colhead{(s)} &  &  & }
\startdata
593.2  &  612.9  & $UVB$ &  17.09  &  0.13 \\
767.0  &  786.7  & $UVB$ &  17.21  &  0.14 \\
1147.9  &  1332.8  & $UVB$ &  17.50  &  0.13 \\
40736.7  &  41644.4  & $UVB$ &  20.21  &  0.14 \\
56651.5  &  57558.5  & $UVB$ &  20.72  &  0.26 \\
337.2  &  586.9  & $UVU$ &  17.67  &  0.08 \\
742.1  &  761.9  & $UVU$ &  17.78  &  0.25 \\
1122.9  &  1317.9  & $UVU$ &  17.77  &  0.18 \\
28831.1  &  29661.9  & $UVU$ &  20.3  &  0.21 \\
46079.7  &  46907.8  & $UVU$ &  20.61  &  0.27 \\
92151.0  &  92979.8  & $UVU$ &  20.83  &  0.31 \\
668.4  &  688.3  & $UVV$ &  15.56  &  0.11 \\
841.2  &  861.0  & $UVV$ &  15.55  &  0.11 \\
1049.4  &  1243.1  & $UVV$ &  15.80  &  0.09 \\
34595.2  &  35419.7  & $UVV$ &  18.07  &  0.07 \\
51699.8  &  52523.8  & $UVV$ &  18.64  &  0.10 \\
80607.5  &  81514.3  & $UVV$ &  19.10  &  0.13 \\
97761.4  &  98526.1  & $UVV$ &  19.54  &  0.20 \\
\enddata
\tablecomments{{$^{\rm{a}}$Since \swift/BAT trigger (add 3199~s to convert to time relative to \fermi/GBM trigger, and see also \citealt{wkd+23})}. Magnitudes are in the native \swift/UVOT system and have not been corrected for extinction in the Milky Way or GRB host galaxy.}
\end{deluxetable}

\subsection{X-rays}
\subsubsection{\nustar\, (3-79 keV)}
\label{text:data:nustar}
The Nuclear Spectroscopic Telescope Array (\nustar, \citealt{hcc+13}) acquired a first set of four observations between 1.96 -- 23.73 days (PIs Margutti \& Racusin) with exposure times of $\approx 20$\,ks, followed by one deeper exposure at 31.91 days (exposure time of $40$\,ks, PI Troja). We reduced the \nustar\, data following standard procedures with the \nustar\, Data Analysis Software (NuSTARDAS) version 0.4.9 and \nustar\, CALDB version 20221130, applying standard filtering criteria with \texttt{nupipeline}. \grb\, is a bright source of hard X-rays in the \nustar\, bandpass (3-79 keV) at all times. For each epoch we extracted a spectrum with \texttt{nuproducts} using a source extraction region centered at the location of the radio counterpart and different sizes to maximize the S/N as reported in Table \ref{Tab:NuSTAR}. We used a source-free background region of radius $>2'$. We find that the hard X-ray spectrum is well modeled by a simple power-law with photon index\footnote{We define the photon index with the convention, $F_{\rm E}\propto E^{-\Gamma}$.} of $\Gamma\approx 1.86$ at all times.

\begin{deluxetable*}{ccccccc}
\label{Tab:NuSTAR}
\tablecaption{Log of \nustar\,observations}
\tablehead{
\colhead{Start Date/Time} & Stop Date/Time&\colhead{Centroid MJD} & \colhead{Time$^{\rm{a}}$} & \colhead{Net Exposure$^{\rm{b}}$} & \colhead{Net Exposure$^{\rm{c}}$} & \colhead{Source region size}  \\
(dd-mm-yy/hh:mm:ss) & (dd-mm-yy/hh:mm:ss)   & (d) & (d) & A (ks) &  B (ks)
& Radius (')}
\startdata
2022-10-11/10:04:09 & 2022-10-11/14:45:00 & 1.96 & 1.96 & 20.66 & 20.49 & 1.5 \\
2022-10-15/05:21:09 & 2022-10-15/17:16:09 & 5.92 & 5.92 & 20.66 & 20.49 & 1.1  \\
2022-10-20/01:06:09 & 2022-10-20/11:56:09 & 10.72 & 10.72 & 20.44 & 20.26 & 1.0 \\
2022-11-02/06:06:09 & 2022-11-02/17:01:09 & 23.93 & 23.73 & 21.30 & 21.09 & 1.0  \\
2022-11-09/23:06:09 & 2022-11-10/23:01:09 & 31.91 & 31.91 & 40.78 & 40.38 & 1.0  \\
\enddata
\tablecomments{$^{\rm{a}}$ With respect to the \fermi/GBM trigger time. \\ $^{\rm{b}}$ For NuSTAR module A. \\ $^{\rm{c}}$ For NuSTAR module B. }
\end{deluxetable*}

\subsubsection{\swift/XRT (0.3-10 keV)}
\label{text:data:XRT}
The \swift\ X-ray Telescope (XRT) began observing \grb\ $\approx3.4$~ks after the \swift/BAT trigger and $\approx6.6$~ks after the \fermi/GBM trigger. We extracted XRT PC-mode spectra at the times corresponding to the \nustar\ epochs using the time-sliced spectrum tool on the \swift\ website\footnote{\url{https://www.swift.ac.uk/xrt_spectra/01126853/}} and modeled the spectra, together with corresponding calibration files, in \texttt{XSPECv12.12.1}. While the derived photon index appears to increase with time (from $\approx1.55$ to $\approx1.82$, tying $N_{\rm H,int}$ across epochs), the evidence for this is marginal ($\lesssim2\sigma$) and we do not consider this statistically significant (but see also \citealt{wkd+23}). Tying the photon index across epochs gives results consistent with the parameters on the \swift\ website. 

Finally, we combine the XRT and \nustar\ data together and perform three joint spectral fits: (i) tying both $N_{\rm H,int}$ and $\Gamma_{\rm X}$ across epochs; (ii) tying $N_{\rm H,int}$ and allowing $\Gamma_{\rm X}$ to vary; and (iii) tying $\Gamma_{\rm X}$ and allowing $N_{\rm H,int}$ to vary. We do not find evidence for evolution in $\Gamma_{\rm X}$ with time. There is marginal evidence in these fits for a decrease in $N_{\rm H,int}$ from $1.51^{+0.13}_{-0.05}\times 10^{22}\,\rm{cm^{-2}}$ to $0.99^{+0.31}_{-0.29}\times 10^{22}\,\rm{cm^{-2}}$. However, these numbers are consistent at the $2\sigma$ level, and we do not consider varying $N_{\rm H,int}$ further. On tying both quantities across epochs, we find $N_{\rm H,int}=1.35^{+0.06}_{-0.09}\times 10^{22}\,\rm{cm^{-2}}$ and $\Gamma_{\rm X}=1.8566^{+0.0033}_{-0.0063}$. 

We download the XRT 0.3–10 keV count rate light curve \footnote{Obtained from the Swift website at \url{http://www.swift.ac.uk/xrt_curves/01126853} and rebinned to a minimum signal-to-noise ratio per bin of 10.} \citep{ebp+07,ebp+09} and convert it to flux density at 1 keV for subsequent analysis (after shifting the time to the GBM trigger time) using $\Gamma_{\rm X}=1.8566$. For this we use unabsorbed counts-to-flux conversion rates of $1.00\times10^{-10}\,{\rm erg\,cm^{-2}\,ct^{-1}}$ and $1.07\times10^{-10}\,{\rm erg\,cm^{-2}\,ct^{-1}}$ for the WT and PC-mode, respectively, obtained from the \swift\ website. We also extract the \nustar\ flux in the range 15--20\,keV and convert it to flux density at 15\,keV using the photon index from the joint fit for subsequent analysis.

\subsection{$\gamma$-rays: \fermi/LAT (100~MeV--100~GeV)}
The LAT instrument on board the Fermi satellite is sensitive to gamma-ray photons in the energy band from 30 MeV to 300 GeV \citep{aaa+09}. We extracted and analyzed the Fermi/LAT data of GRB 221009A using the public software {\sc gtburst}, which is distributed as part of the official Fermitools software package\footnote{\url{https://fermi.gsfc.nasa.gov/ssc/data/analysis/scitools/gtburst.html}}. We extracted the LAT data within a temporal window from 3.5~ks to 100~ks (0.04--1.16~days) after the GBM trigger time. We filtered photons with energies in the 100~MeV--100~GeV range, within a region of interest (ROI) of 12$^{\circ}$ centred on the burst position of R.A.~= 288.264$^{\circ}$ and Dec.~= 19.773$^{\circ}$, and with an angle from the spacecraft zenith $<100^{\circ}$, as part of the standard procedure. We selected the ``P8R3 SOURCE'' class as the instrument response function, suitable for late-time emission subsequent to the prompt phase of the burst. We extracted light curves assuming a power-law model for the spectrum of the source, together with the ``isotropic template'' and ``template'' for the particle background and the Galactic component, respectively. We performed an unbinned likelihood analysis, setting the minimum test statistics (TS) to 10 for detections. We calculated flux upper limits using a spectral slope of $-2$ in the energy band of 100 MeV–100 GeV. We report our \fermi/LAT photometry in Table~\ref{tab:data:LAT}. 

\begin{deluxetable*}{ccccccccc}
\label{tab:data:LAT}
\tablecaption{\fermi/LAT Observations}
\tablehead{
\colhead{Start Time$^{\rm{a}}$} & \colhead{Stop Time$^{\rm{a}}$} & \colhead{Energy Flux} & \colhead{Energy Flux} & \colhead{Photon Flux} & \colhead{Photon Flux} & \colhead{Photon Index$^{\rm{b}}$}  & \colhead{Photon Index} & \colhead{Test}\\
    &     &                            &  Uncertainty               & & Uncertainty & &  Uncertainty & Statistic \\
(s) & (s) & (erg\,s$^{-1}$\,cm$^{-2}$) & (erg\,s$^{-1}$\,cm$^{-2}$) & (ph\,s$^{-1}$\,cm$^{-2}$) & (ph\,s$^{-1}$\,cm$^{-2}$) & \\
}
\startdata
3000.0  & 4429.2 & $2.67\times10^{-08}$ & $1.43\times10^{-08}$ & $3.06\times10^{-05}$ &        $1.06\times10^{-05}$ & 2.10 & 0.283 & 57 \\
4429.2  & 6539.4 & $9.54\times10^{-09}$ & $2.69\times10^{-09}$ & $1.12\times10^{-05}$ &        $1.88\times10^{-06}$ & 2.11 & 0.141 & 177 \\
9654.9  & 14254.6 & $4.44\times10^{-09}$ & $1.73\times10^{-09}$ & $5.41\times10^{-06}$ & 
  $1.37\times10^{-06}$ & $2.13$ & 0.209 & 79 \\
14254.6 & 21045.8 & $1.72\times10^{-09}$ & $9.79\times10^{-10}$ & $2.35\times10^{-06}$ & 
  $9.06\times10^{-07}$ & $2.19$ & 0.334 & 30 \\
21045.8 & 31072.3 & $1.44\times10^{-09}$ & $7.95\times10^{-10}$ & $2.01\times10^{-06}$ &       $8.81\times10^{-07}$ & $2.20$ & 0.354 & 21 \\
31072.3 & 45875.7 & $5.70\times10^{-10}$ & $2.79\times10^{-10}$ & $1.17\times10^{-06}$ &       $5.26\times10^{-07}$ & $2.46$ & 0.402 & 13 \\
45875.7 & 67731.6 & $4.34\times10^{-10}$ & $1.80\times10^{-10}$ &  $1.27\times10^{-06}$ &      $5.08\times10^{-07}$ & $2.87$ & 0.445 & 12 \\
67731.6 & 100000. & $<6.42\times10^{-10}$ & \ldots &  $<5.79\times10^{-07}$ & 
   \ldots              & $2.00$ &\ldots & 5 \\
\enddata
\tablecomments{$^{\rm{a}}$ With respect to the \fermi/GBM trigger time. $^{\rm{b}}$ Here defined as $F_{\rm E}\propto E^{-\Gamma}$.}
\end{deluxetable*}

\section{Multi-wavelength modeling}
\begin{figure*}
    \centering
    \begin{tabular}{cc}
        \includegraphics[width=0.48\textwidth]{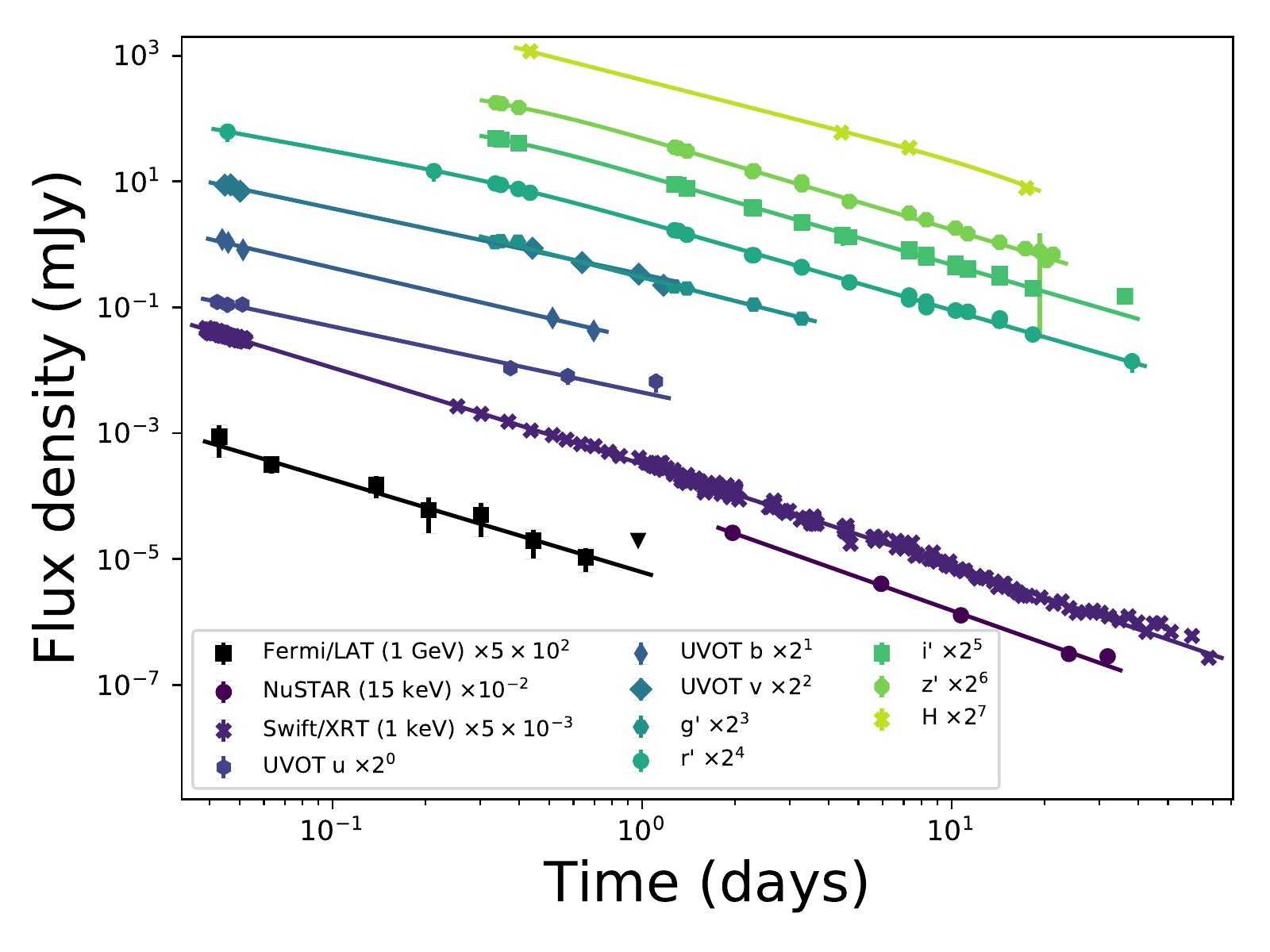} &
        \includegraphics[width=0.48\textwidth]{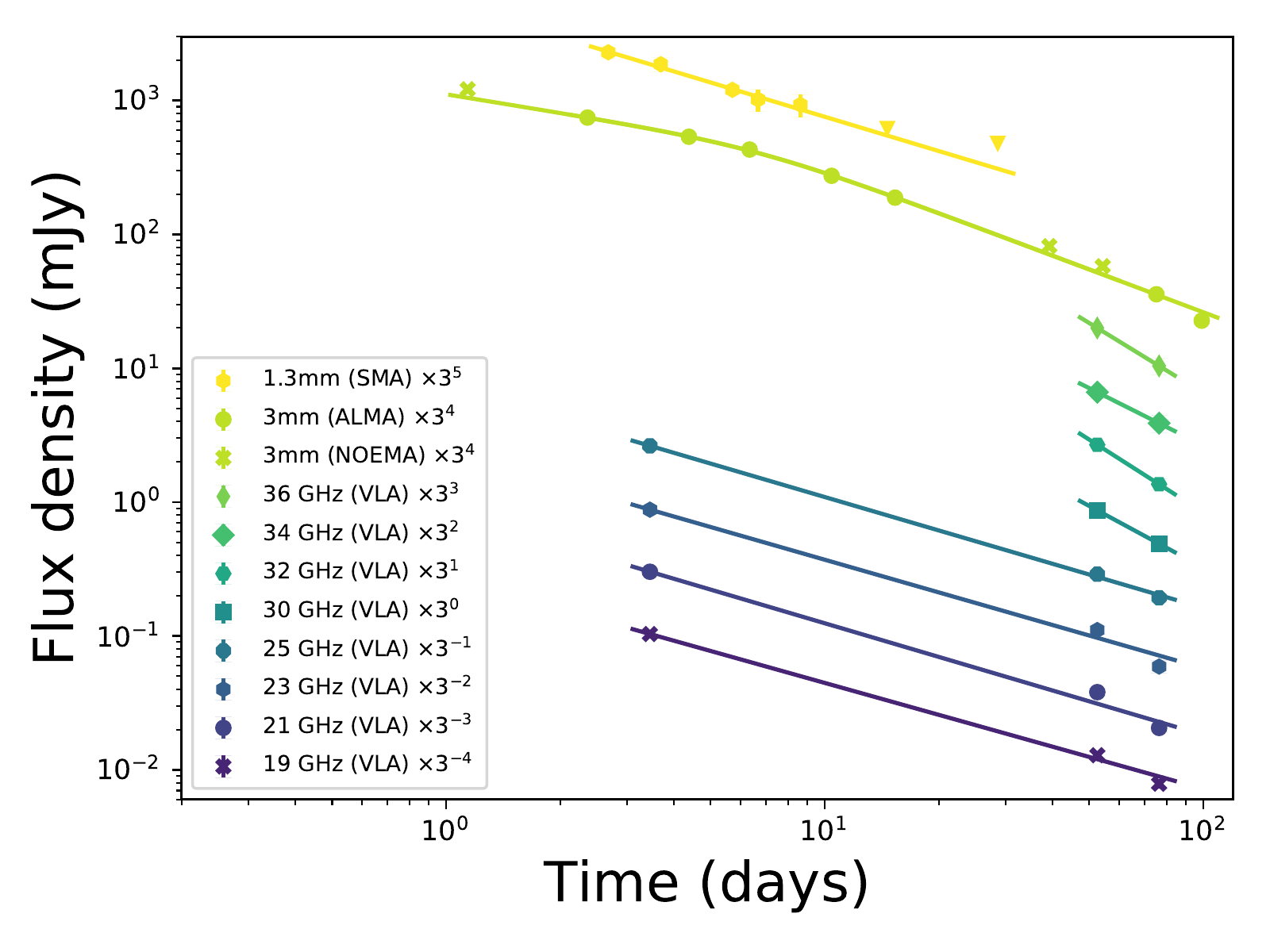} \\
        \includegraphics[width=0.48\textwidth]{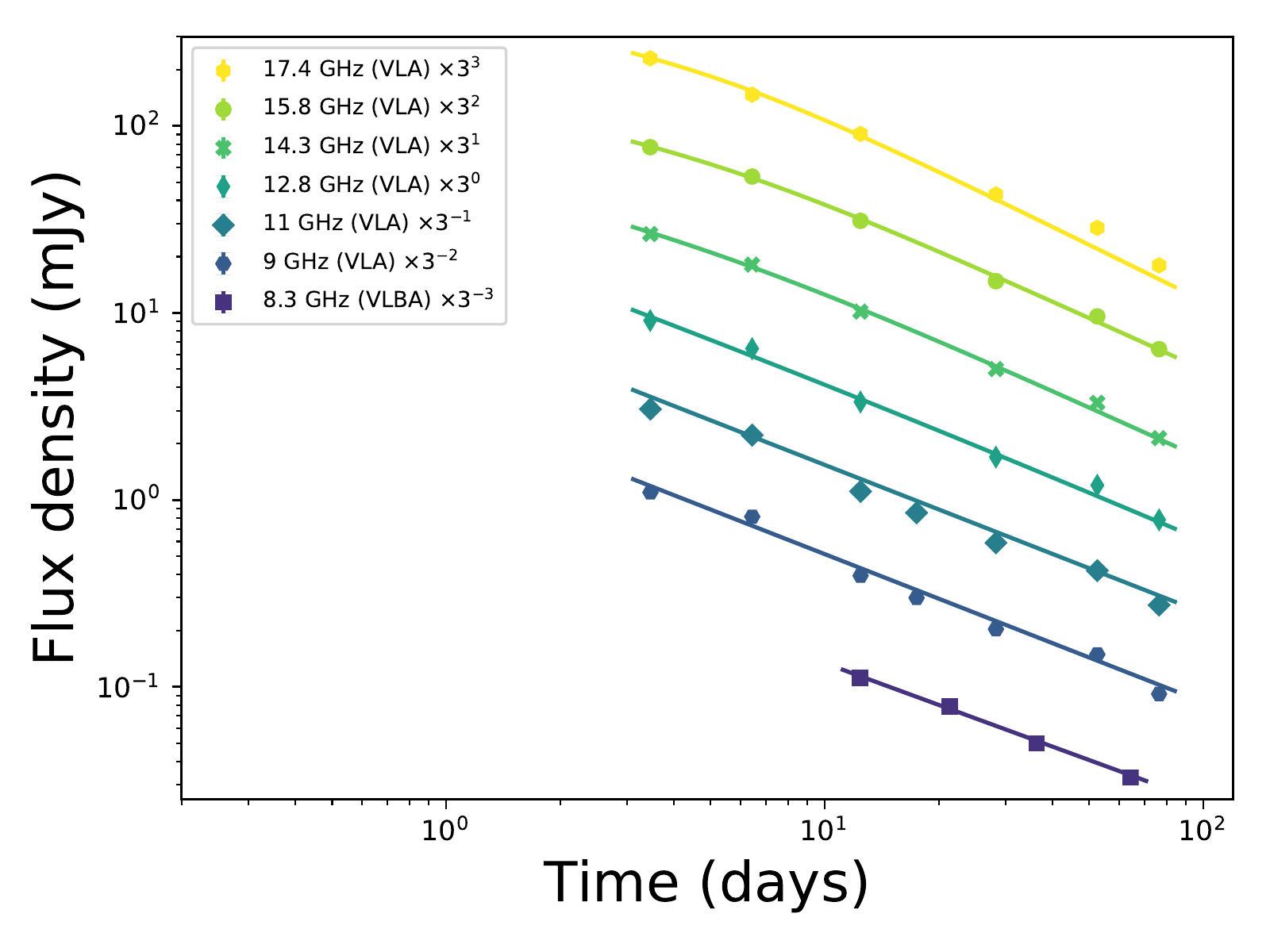} &
        \includegraphics[width=0.48\textwidth]{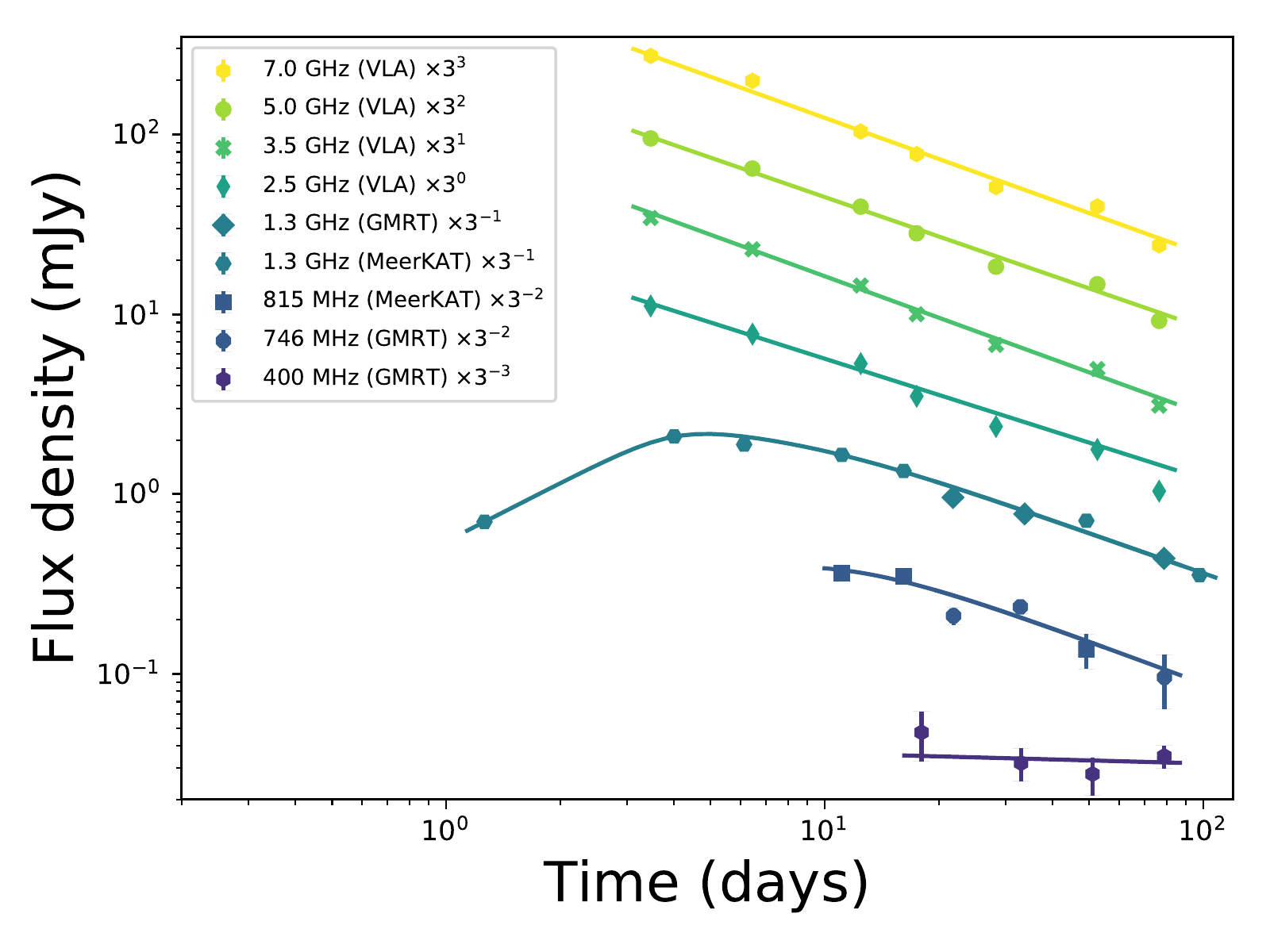}
    \end{tabular}
    \caption{Light curves of \grb\ from 1~GeV to 400~MHz together with the broken power-law model fits used to interpolate data to a common time and to inform the multi-wavelength modeling (Sections~\ref{text:fs} and \ref{text:radiocomponents}). The light curves have been multiplied by the factors listed in the legends to offset them in the plot for clarity. The three panels with radio data are on the same temporal scale to facilitate comparison.}
    \label{fig:lcbplfits}
\end{figure*}

We now interpret the multi-wavelength afterglow observations in the context of the standard model of synchrotron radiation produced by a relativistic forward shock (FS) produced by the GRB jet propagating into the pre-explosion environment. The model is parameterized by the shock energy, $\EKiso$, the radial density profile $\rho\propto r^{-k}$ of the environment, the density normalization ($\dens$ for $k=0$, ISM-like; and $\Astar$ for $k=2$, wind-like\footnote{$\Astar=1$ corresponds to a mass-loss rate of $\dot{M_w}=10^{-5}M_\odot/{\rm yr}$ for a wind velocity of $v_w=10^3\,{\rm km\,s}^{-1}$.}), the fraction of shock energy in relativistic electrons (\epse) and magnetic fields (\epsb), as well as the index ($p$) of electrons accelerated to a power-law energy distribution. The observed radiation is expected to be characterized by three break frequencies: (i) the self-absorption frequency, $\nua$; (ii) the characteristic frequency, $\numax$; and (iii) the cooling break, $\nuc$. Solving for these parameters requires interrogating the observed light curves and spectral energy distributions (SEDs). 

\subsection{X-ray and optical/NIR: Forward Shock}
\label{text:fs}

In order to interpolate observations to common times for constructing and plotting SEDs, we fit the observed multi-wavelength light curves to a series of broken power law models. The resulting fits are presented in Figure~\ref{fig:lcbplfits}. 
The \fermi/LAT 1~GeV and \nustar\ 15~keV light curves can be fit with a single power law with decay indices\footnote{We use the convention $F_\nu\propto t^\alpha\nu^\beta$ throughout.}, $\alphaLAT=-1.47\pm0.08$ and $\alpha_{\rm 15\,keV}=1.74\pm0.03$. The \swift/XRT 1~keV light curve can be fit with a broken power law with decay indices, $\alpha_{\rm X,1}=-1.46\pm0.02$ and $\alpha_{\rm X,2}=-1.70\pm0.03$; however, the break time, $t_{\rm b,X}=(1.0\pm0.8)$~day is not well constrained. The $r'$-band light curve can be fit with a broken power law with decay indices, {$\alpha_{\rm r,1}=-0.92\pm0.10$ and $\alpha_{\rm r,2}=-1.42\pm0.01$ and break time, $t_{\rm b,r}=(0.40\pm0.04)$~days}, although the pre-break decay is contingent upon uncertain photometry reported in GCN circulars. Correcting the $r'$-band flux for Milky Way extinction, the NIR-to-optical spectral index at $\approx4.4$ days is $\betaniropt=-0.76\pm0.04$, while the NIR-to-X-ray spectral index is similar, $\betanirx=-0.70\pm0.01$. Spectral indices involving NIR/optical bands are subject to additional, indeterminate systematic errors due to unknown uncertainties on the Galactic extinction along the line of sight and any additional intrinsic extinction. 

\begin{figure*}
    \centering
    \begin{tabular}{cc}
        \includegraphics[width=0.48\textwidth]{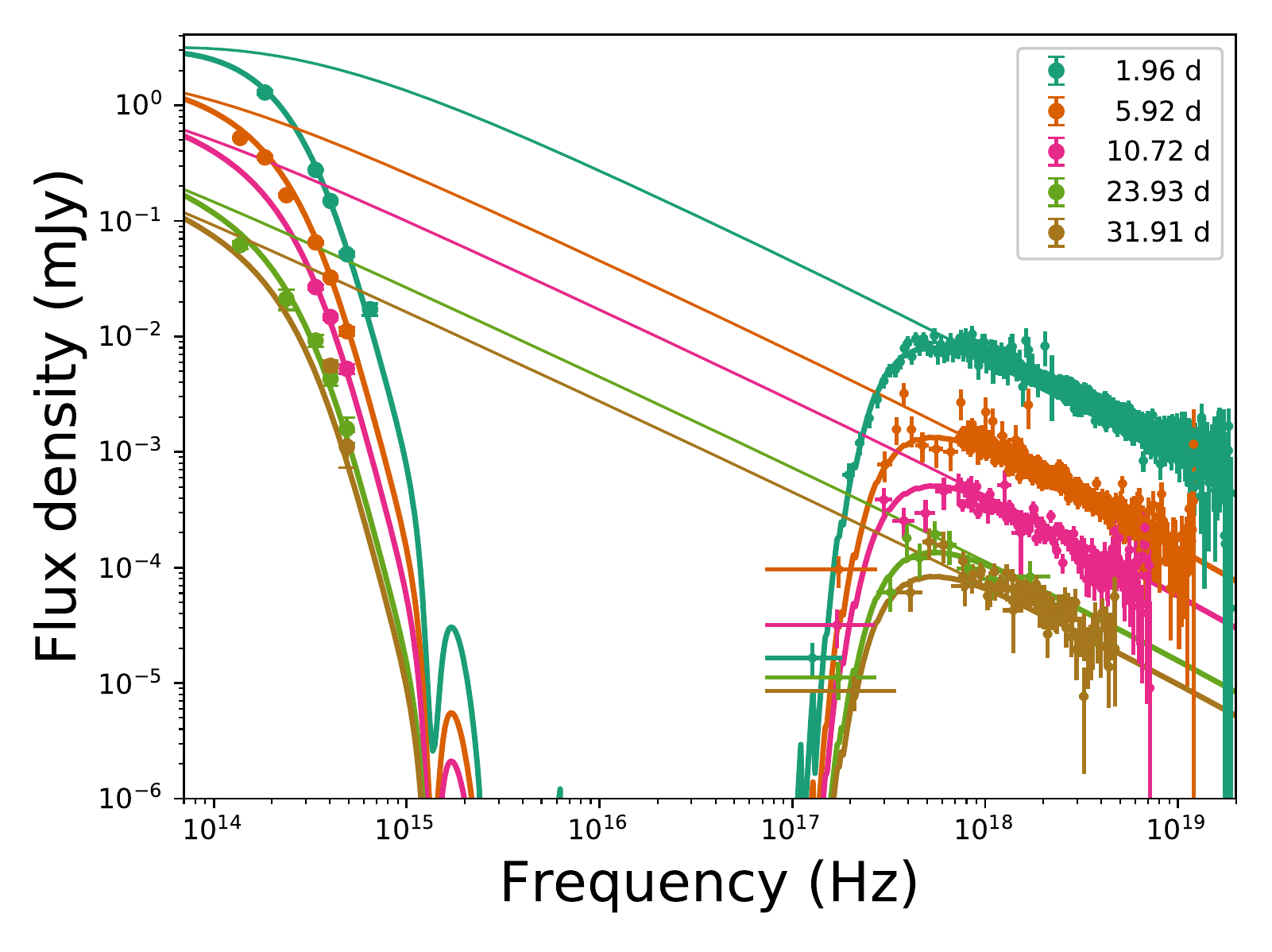} &
        \includegraphics[width=0.48\textwidth]{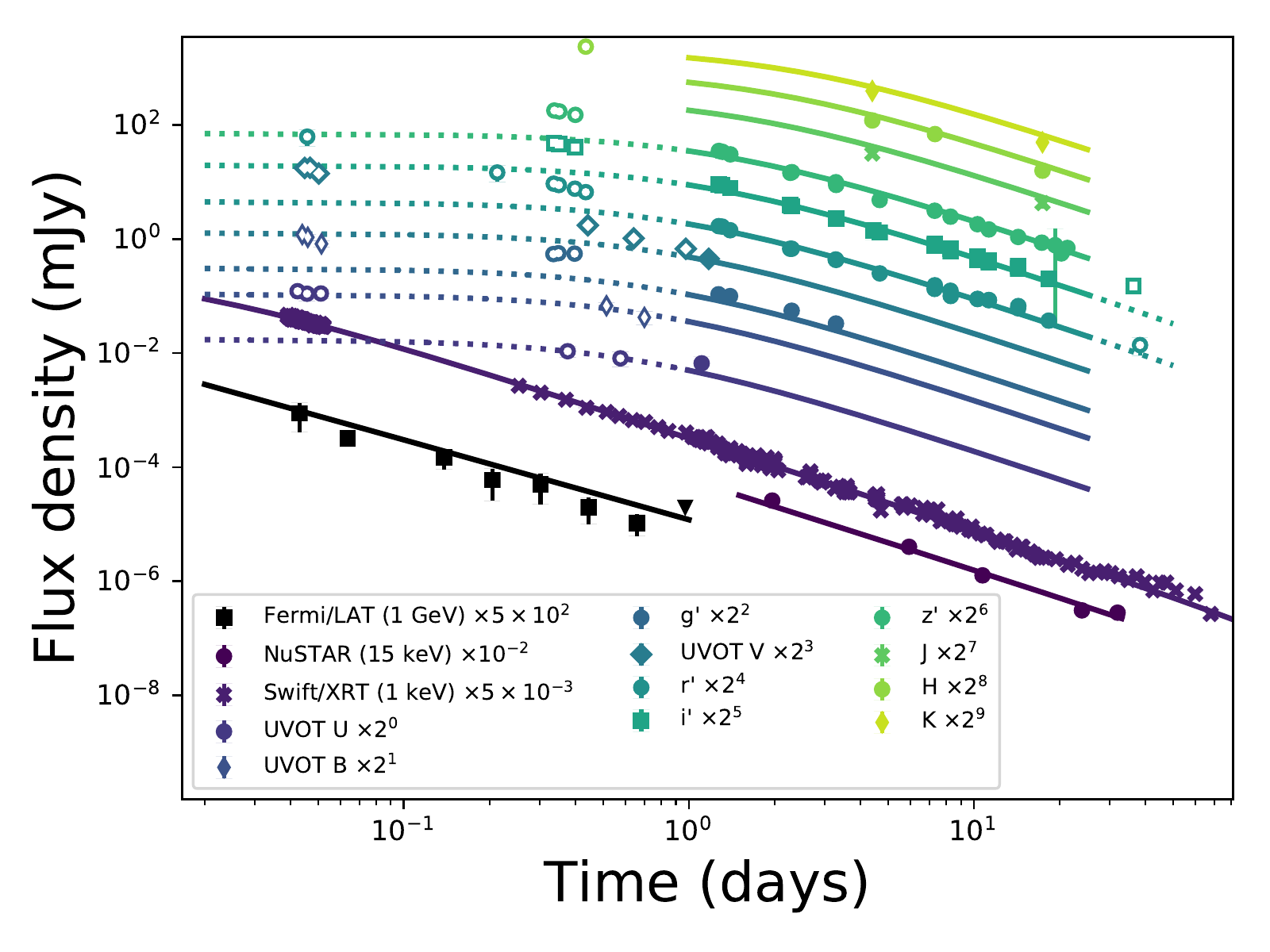} 
    \end{tabular}
    \caption{SEDs (left) and a sampling of available light curves (right) at X-ray, UV, optical, and NIR wavelengths of \grb\ (data points), together with the highest-likelihood FS wind model (lines) that explains most of these observations (Section~\ref{text:fs}). Thin lines indicate the underlying synchrotron spectrum. The break at $\approx2\times10^{14}$~Hz in the first epoch is$\numax$. The X-ray SEDs shown here are the result of joint spectral fits of \swift\ and \nustar\ observations at the corresponding times (Section~\ref{text:data:XRT}). Optical points have been interpolated using the best-fit broken power-law light curves (Figure~\ref{fig:lcbplfits}). The model cannot explain the UV/optical light curves at $\lesssim1$~day; these points (open symbols) were excluded from the fit. These caveats are discussed further in Section~\ref{text:fscaveats}.}
    \label{fig:fsfits-NuSTAR}
\end{figure*}

A shallower light curve in the optical compared to the X-rays usually indicates\footnote{Whereas the presence of a SN could also make optical light curves shallower, as we discuss later in this section, we find no strong additional evidence for such a component.} an ISM-like environment with $\nuopt<\nuc<\nux$. However, a uniform-density environment is ruled out for this burst by the closure relations between the light curve decay rates and spectral indices. To see this, we start with the optical spectral index of $\betaniropt\approx-0.76$, which requires an electron energy index of $p\approx2.52$ in the regime $\nuopt<\nuc<\nux$. This would imply $\alphax=(2-3p)/4 \approx -1.4$, which is shallower than observed. Furthermore, the similarity between $\betaniropt$ and $\betanirx$ suggests that no spectral break is present between these bands. If we instead consider the regime $\nuopt<\nux<\nuc$, and $p\approx2.52$ from the spectral indices, then the predicted light curve decline rates of $\alphaopt\approx\alphax\approx-1.1$ are significantly shallower than observed in either band. Finally, in the regime $\nuc<\nuopt<\nux$, the observed spectral indices would indicate $p\approx1.5$, which predicts a light curve decline rate of $\alpha=-(3p+10)/16\approx-0.9$ \citep{dc01,zm04}, again much shallower than observed in either band. While this can be remedied by interposing a jet break (after which the light curve declines at $\approx-p$, including exponential sideways spreading; \citealt{rho99,sph99}) at $\tjet\approx1$~day, in this model the jet quickly becomes non-relativistic and the resulting evolution cannot match the radio observations (Section~\ref{text:radiocomponents}). We note that \cite{smoy22} came to a similar conclusion even without radio data, but incorporating \fermi/LAT GeV observations. For completeness, we present an ISM model, together with its shortcomings, in Appendix~\ref{appendix:ISM}.

We next consider the other standard case of $k=2$ for the density profile of the pre-explosion environment. The $\nuc<\nuopt<\nux$ regime is ruled out as in the ISM case, as in this regime the light curves are agnostic to the density profile. In the regime $\nuopt<\nuc<\nux$, the optical spectral index again requires $p\approx2.52$. The predicted light curve decay rates are $\alphaopt\approx-1.64$ and $\alphax\approx-1.39$; however, the opposite is observed\footnote{This remains a fundamental issue in modeling this burst, as we discuss in Section~\ref{text:fscaveats}.}. In addition, in this regime we expect an X-ray spectral index of $\betax=-p/2\approx-1.26$, which is significantly steeper than the observed. Finally, we consider the regime $\nuopt<\nux<\nuc$. The spectral index between the \swift/XRT band at 1~keV and the \fermi/LAT band at 1~GeV is $\beta_{\rm keV-GeV}=-1.10\pm0.04$. This is significantly steeper than the X-ray spectral index alone. Furthermore, the spectral index in the GeV band of $\beta_{\rm GeV}=1-\Gamma_{\rm GeV}\approx-1.2$ (using a weighted-mean photon index, Table~\ref{tab:data:LAT}) also appears steeper than the X-ray data alone, supporting the presence of a spectral break between the X-ray and GeV bands. While the optical and X-ray spectral indices and decay rates ought to be equal to each other in this regime, a potential solution\footnote{\cite{fsr+23} assume the same spectral regime of $\nuopt<\nux<\nuc$ and instead explain the shallower optical light curve compared to the X-rays as contribution from an underlying SN in the optical. However, they also require strongly time-varying extinction for this scenario to explain the observed optical color.} for the shallower optical decay observed could be the proximity of $\numax$ to the optical/NIR bands.

We find that we are able to find a relatively satisfactory solution with $k=2$ for $p\approx2.5$ in the regime $\numax\lesssim\nuopt<\nux<\nuc<\nu_{\rm LAT}$ (Figure~\ref{fig:fsfits-NuSTAR}) with the following caveats:
(i) this model does not match any of the radio observations (although this is also true for the ISM model, see Appendix~\ref{appendix:ISM}), and over-predicts the  ALMA 97.5~GHz data point at $\approx100$~days; 
(ii) the model over-predicts the NIR ($JHK$) flux at the $\approx30\%$ level; and 
(iii) in the optical, this model under-predicts the observed light curves at both $\lesssim1$~day and $\gtrsim30$~days. 
We discuss point (i) in Section~\ref{text:radiocomponents} and 
points (ii) and (iii) in Section~\ref{text:fscaveats}. 

To explore this FS model further, we fit the \fermi/LAT light curve at 1~GeV, the \swift/XRT light curve at 1~keV, the UV/optical/NIR data at $1$--30~days, and the 97.5~GHz data point at $\approx100$~days simultaneously by sampling the parameter space of $p$, \EKiso, \Astar, \epse, \epsb, and \tjet\ using Markov Chain Monte Carlo with \texttt{emcee} \citep{fhlg13}. The details of our implementation are described in \citet{lbz+13,lbt+14}. We run 512 walkers for 5k steps and discard the first 100 steps as burn-in. We use a uniform prior on $p$ from 2.001 to 2.99. We restrict $\log(\epse)$ and $\log(\epsb)$ to the range $\in(-5,0)$ and require $\epse+\epsb<1$. We constrain $\log(\Astar)\in(-10,10)$,  $\EKiso/{\rm erg}\in(10^{48},5\times10^{58})$, and $\log(\tjet/{\rm days})\in(-5,5)$ and use \cite{jef46} priors for these last five parameters. To account for systematic flux calibration offsets in any given band in our data set as well as to prevent single, high signal-to-noise points from driving the entire fit, we implement a minimum uncertainty floor of 10\% prior to running the fit. 
We fix the Galactic extinction to $A_{\rm V,gal}=4.1$~mag, and find some evidence for additional extinction, $A_{\rm V,host}\approx0.2$~mag. However, we note that there is a degeneracy between $A_{\rm V,gal}$ and $A_{\rm V,host}$ as the redshift is low; thus it is entirely possible that the true Galactic extinction is lower and the true extinction along the line of sight through the host galaxy is higher than the inferred value. 

In the highest-likelihood model, the synchrotron break frequencies are located at $\nua\approx2\times10^5$~Hz, $\numax\approx4.5\times10^{14}$~Hz, and $\nuc\approx3.8\times10^{19}$~Hz at 1~day, with $\numax\approx\nuopt<\nux<\nuc<\nu_{\rm LAT}$, as expected. The proximity of $\numax$ to the optical contributes to a shallower optical light curve than expected for the regime $\numax<\nuopt<\nuc$. The proximity of $\nuc$ to the X-rays results in a spectral index intermediate between $\beta=(1-p)/2\approx-0.76$ and $\beta=-p/2\approx-1.3$. The resulting SED fits the \nustar\ spectra well, even though those data were not included in the fit. Finally, the spectral index of $\beta=-p/2\approx-1.3$ closely corresponds to the value of $\beta_{\rm GeV}\approx-1.2$ in the \fermi/LAT band at 1~GeV. {The proximity of $\nuc$ to the X-rays contributes to curvature in the \swift/XRT light curve, and this curvature is sufficient to explain the ``break'' inferred from a broken power-law fit to the data at 1\,keV (Figure~\ref{fig:lcbplfits}). Thus a jet break at $\approx1$~day \citep{gcn32755} is not required by the data under this model.}

\begin{figure*}
  \centering
  \includegraphics[width=0.9\textwidth]{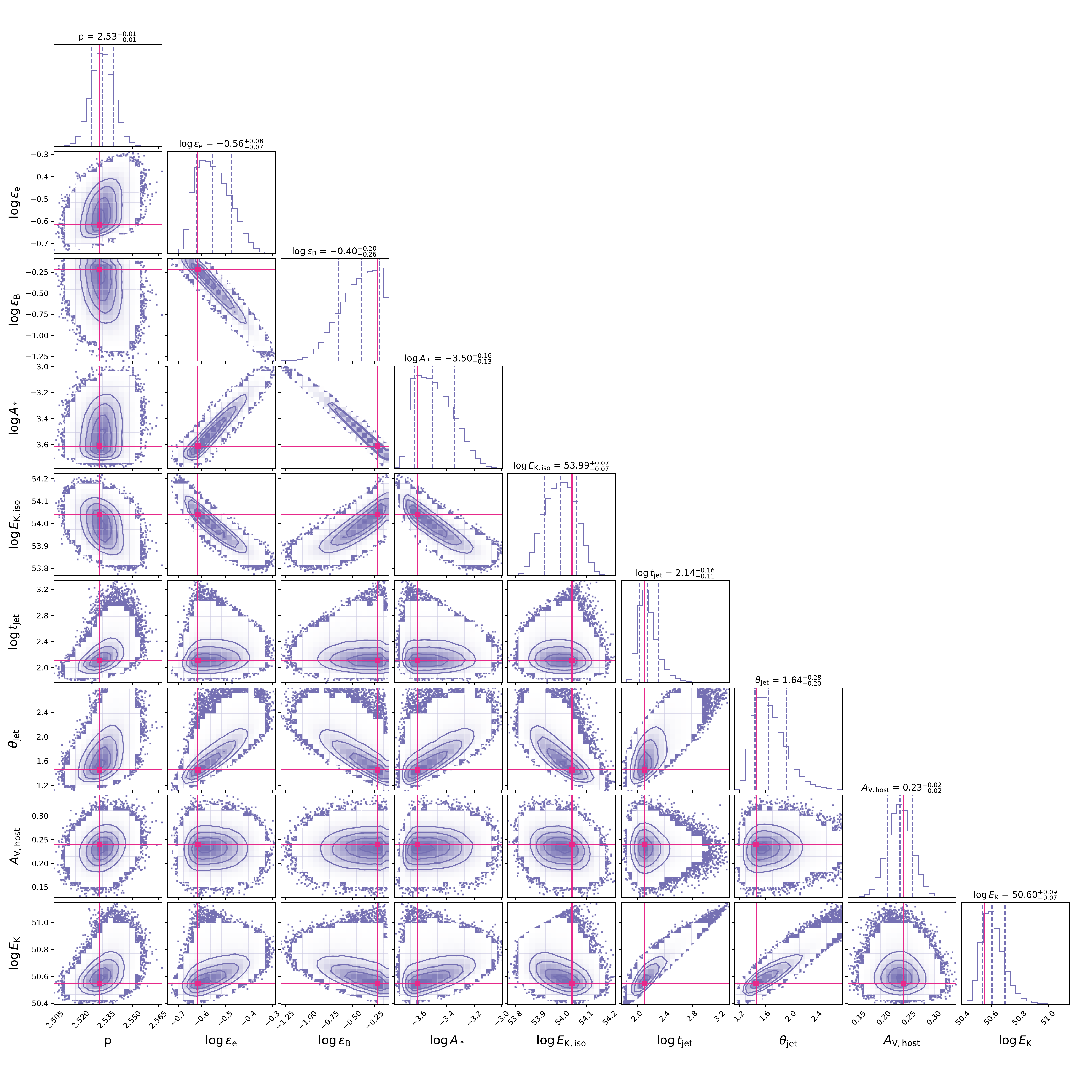}
  \vspace{-0.2in}
  \caption{Correlations and marginalized posterior density from multi-wavelength modeling of \grb\ with \EKiso\ and \EK\ in erg. We compute the opening angle ($\thetajet$) in degrees from \EKiso, \Astar, and \tjet. The contours enclose 39.3\%, 86.4\% and 98.9\% of the probability mass in each correlation plot (corresponding to $1\sigma$, $2\sigma$, and $3\sigma$ regions for two-dimensional Gaussian distributions), while the dashed lines in the histograms indicate the 15.9\%, 50\% (median), and 84.1\% quantiles (corresponding to $\pm1\sigma$ for one-dimensional Gaussian distributions). The distribution for $\epsb$ cuts off at high values due to the prior bound of $\epse+\epsb<1$. Strong correlations are evident between most parameters as $\nua$ is unconstrained in this model. See Table~\ref{tab:params} for a summary.}
\label{fig:corner}
\end{figure*}

We summarize the results of our MCMC analysis in Table~\ref{tab:params}. The electron index $p$ is sharply constrained by the X-ray light curve, the NIR-to-X-ray spectral index, and the X-ray-to-GeV spectral index. As \nua\ is unconstrained in this model, the physical parameters (\EKiso, \Astar, \epse, \epsb) exhibit degeneracies (Figure~\ref{fig:corner}). 
The observed 3mm flux at $\approx100$~days is lower than expected from a spherical FS, and this drives a jet break\footnote{We note that the jet break time is bounded below by the absence of a steep decline in the X-ray light curve to $\tjet\gtrsim70$~days.} at around this time in the model, yielding a small jet opening angle\footnote{{In some models, the jet break is instead interpreted as a viewing angle effect \citep{mrw98,dg01,zm02a}. For a given set of observed light curves, this framework implies a beaming-corrected energy greater than that computed under the standard jet-break interpretation by a factor of $1+2\ln(\thetajet/\theta_c)$, where $\theta_c$ is some narrow opening angle within which the outflow energy per unit solid angle is roughly constant \citep{rlr02}.}}, $\thetajet\lesssim2^\circ$.

As mentioned earlier, this model (i) somewhat ($\approx30\%$) over-predicts the NIR ($JHK$) flux (Figure~\ref{fig:fsfits-NuSTAR}) and (ii) under-predicts the optical emission at $\lesssim1$~day and at $\gtrsim30$~days. We speculate that the mismatch in the NIR may be related to the inability of this model to also fit the radio data, as discussed next in Section~\ref{text:radiocomponents}. We discuss these discrepancies in the NIR and optical, together with additional caveats on the FS modeling, in Section~\ref{text:fscaveats}. 

We note that our model parameters are somewhat different from the analyses of \cite{rwz22} and \cite{smoy22} (notably, we find a much higher value of $\epsilon_B$). However, these papers rely solely on data collected within $\lesssim7$ and $\lesssim12$ d of \grb's discovery, respectively. When compared to our more extensive dataset, we find that both previously proposed models dramatically overpredict the radio emission at $t\gtrsim5$ d and are therefore ruled out. Finally, in contrast to \cite{fsr+23}, we do not include a SN contribution to our model, as this does not appear to be required by the data. In particular, there is no strong evidence for excess emission relative to the afterglow model in any of the optical or NIR light curves. This suggests that the contribution of the SN is lower than the flux level of the observed multi-frequency optical/NIR light curves. Further investigation of the SN requires spectroscopic information, and we defer a detailed discussion of the SN to papers focused on this emission component. We also do not include host galaxy emission in our model. While host emission is known to affect the \textit{HST} data (not included here) at $\approx30$~days \citep{gcn32921}, we do not expect the host galaxy to make a significant contribution in the optical light curves at the earlier times ($\lesssim20$~days) considered here. 

\begin{deluxetable}{lrr}
\tabletypesize{\scriptsize}
\tablecaption{FS Model Parameters \label{tab:params}}
\tablehead{
\colhead{Parameter} & \colhead{Best fit} & \colhead{MCMC\tablenotemark{a}}
}
\startdata
$p$          & $2.531$ & $2.53\pm0.01$\\[2pt]
$\log\epse$  & $-0.616$ & $-0.56^{+0.08}_{-0.07}$\\[2pt]
$\log\epsb$  & $-0.221$ & $-0.40^{+0.20}_{-0.26}$\\[2pt]
$\log\Astar$ & $-3.612$ & $-3.50^{+0.16}_{-0.13}$\\[2pt]
$\log(\EKiso/$erg) & $54.04$ & $53.99\pm0.07$\\[2pt]
$\log(\tjet/$d)      & $2.110$  & $2.14^{+0.16}_{-0.11}$\\[2pt]
$\thetajet$        & $1.46$  & $1.64^{+0.28}_{-0.20}$\\[2pt]
$A_{\rm V,host}$   & $0.24$  & $0.23\pm0.02$\\[2pt]
$\log(\EK/$erg)        & $50.55$ & $50.60^{+0.09}_{-0.07}$\\[2pt]
$\log(\nuaf/$Hz)\tablenotemark{b}    & $5.3$ & \ldots\\[2pt]
$\log(\numf/$Hz)  & $14.6$ & \ldots\\[2pt]
$\log\nucf/$Hz)  & $19.6$ & \ldots\\[2pt]
$\log(\fnumf/$mJy) & $0.98$ & \ldots \\[2pt]
\enddata
\tablecomments{Frequencies and flux densities are calculated at 1~day.
}
\tablenotetext{a}{Summary statistics from the marginalized posterior density distributions, with median and $\pm34.1\%$ quantiles (corresponding to $\pm1\sigma$ for Gaussian distributions; Figure~\ref{fig:corner}).}
\tablenotetext{b}{This frequency is not directly constrained by the data.}
\end{deluxetable}

\subsection{Radio: Multiple Components}
\label{text:radiocomponents}
\begin{figure}
    \centering
        \includegraphics[width=0.48\textwidth]{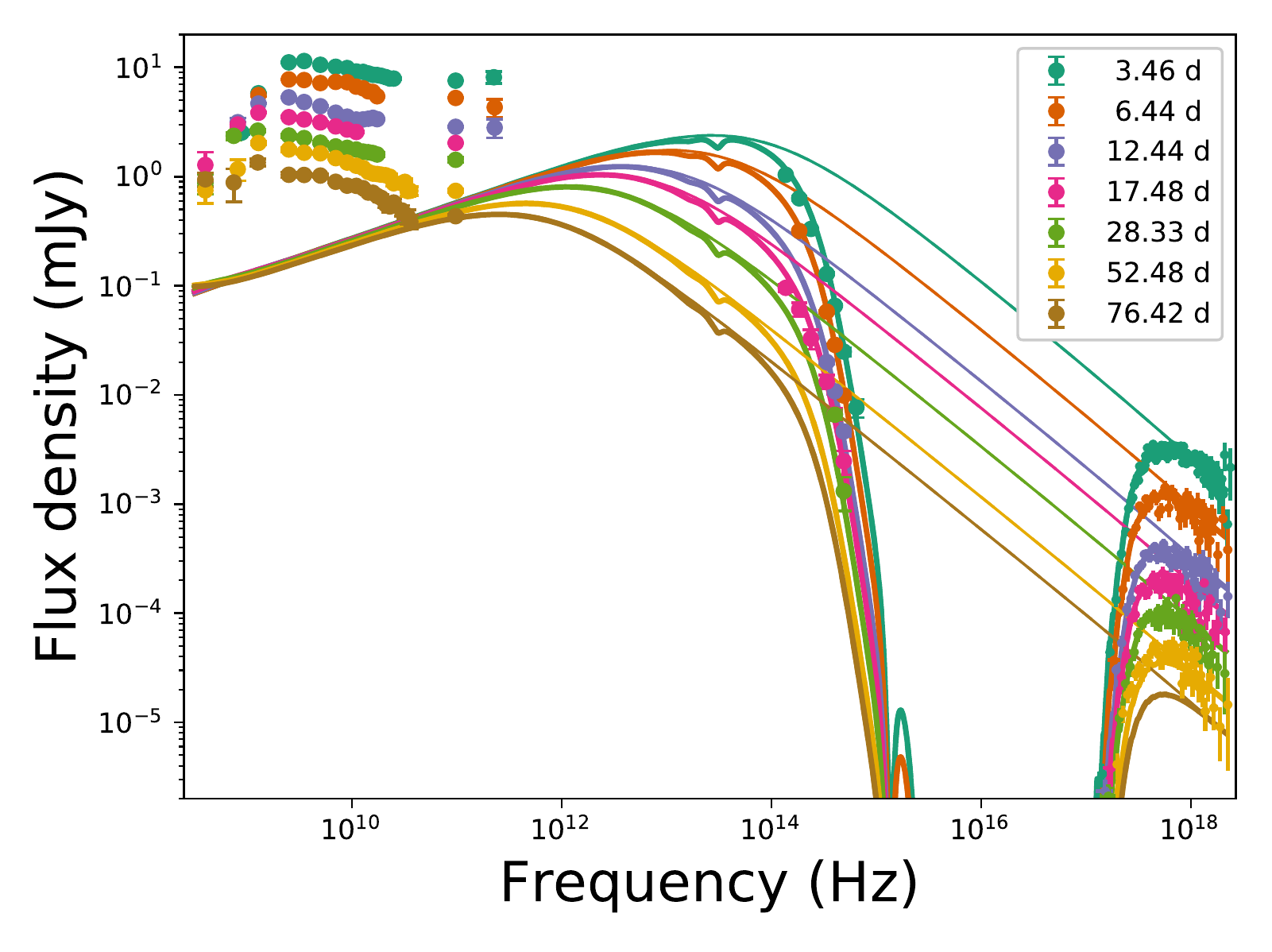}
        \caption{Multi-epoch SEDs from our uGMRT, VLA, ALMA, NOEMA, and SMA programs, compared with optical observations and \swift/XRT SEDs at similar times, together with a wind FS model (lines). The FS cannot explain the observed radio flux and slowly evolving radio peak.}
    \label{fig:windfit}
\end{figure}

In Figure~\ref{fig:windfit}, we present our VLA SEDs together with data from GMRT, ALMA, and NOEMA at 400~MHz, 800~MHz, 1.3~GHz, 97.5~GHz, and 225~GHz interpolated\footnote{The typical temporal dispersion of data points in a given SED is $\approx10\%$.} to the times of the VLA SEDs using their corresponding best-fit broken power-law functions (Figure~\ref{fig:lcbplfits}). We also extract XRT SEDs at the times of the VLA SEDs and fit for the spectral normalization with spectral parameters fixed from the NuSTAR-XRT joint fit (Section~\ref{text:data:XRT}). 

We find that the FS model discussed in Section~\ref{text:fs} under-predicts all the radio observations, except at the highest frequencies (at 97.5~GHz, from ALMA) at $\gtrsim75$~days. The radio emission is strongly self-absorbed below $\approx2$~GHz for most of the period spanning $\approx3$--76~days, while the optically thin spectrum above the radio peak does not match the inverted $\nu^{1/3}$ spectrum expected from the low-energy tail of the minimum energy electrons. In addition, the spectral index above the peak ($\beta\approx-0.2$) is significantly shallower than $\betaniropt$ as well as $\betax$. This indicates that either the radio emission arises from a separate emission component, or the approximation of the electron power-law being truncated at a minimum Lorentz factor ($\gamma_{\rm min}$) breaks down in this case. 

\begin{figure*}
    \centering    
    \begin{tabular}{cc}
        \includegraphics[width=\columnwidth]{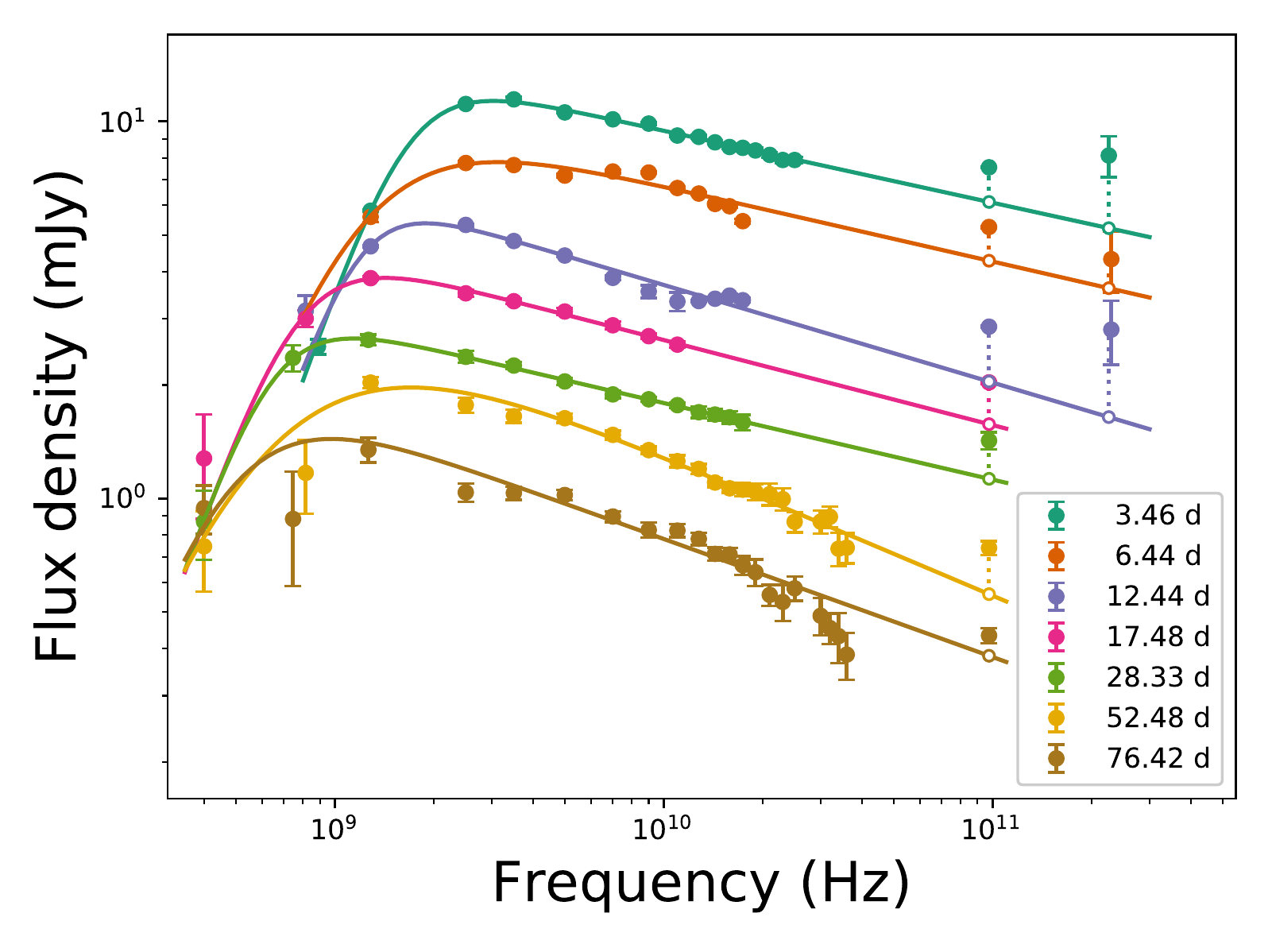} &
        \includegraphics[width=\columnwidth]{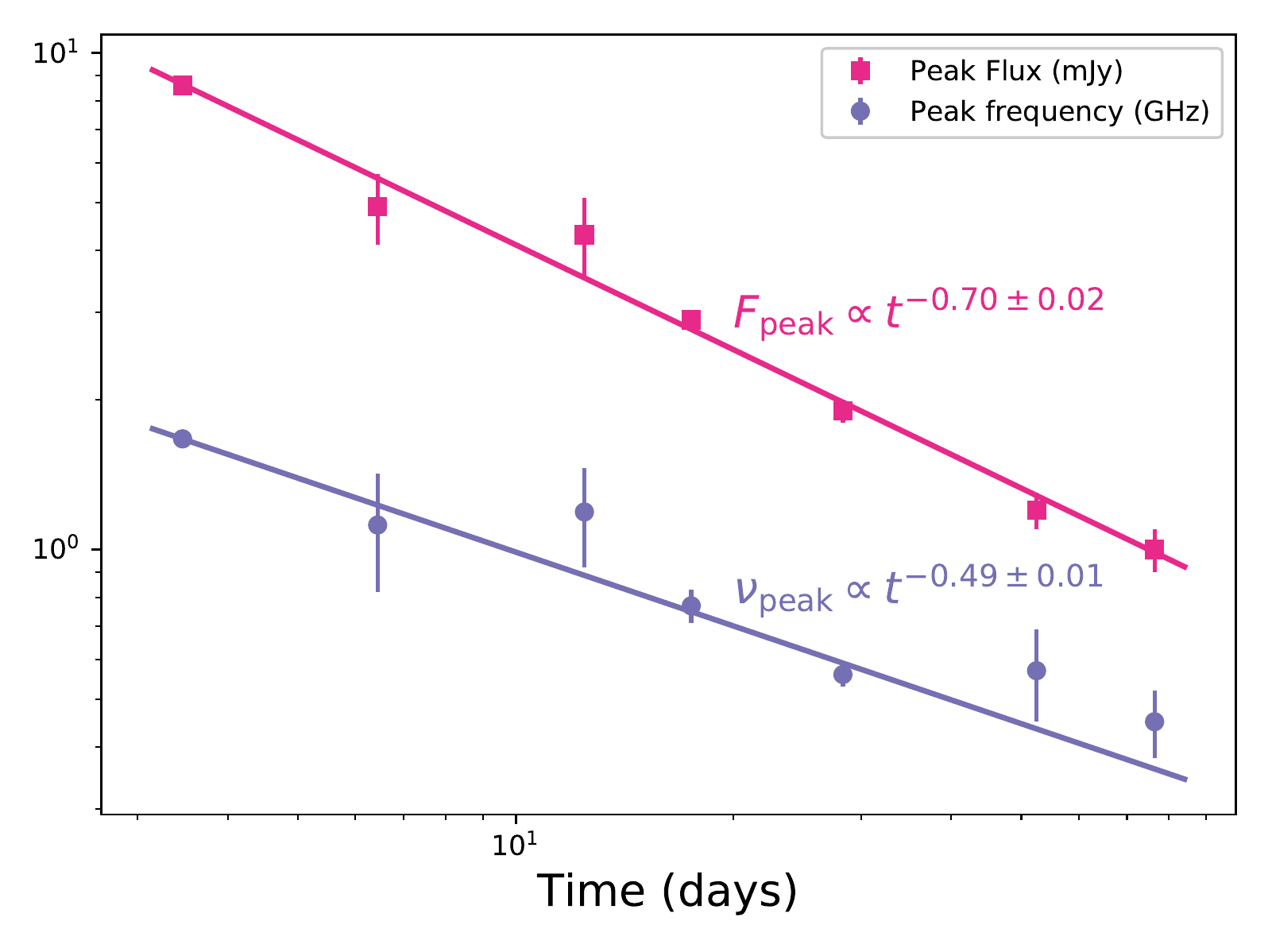}
    \end{tabular}
    \caption{Left: Radio SEDs from our uGMRT, VLA, ALMA, NOEMA, and SMA observations reveal a self-absorbed spectrum with a peak at $\approx2$~GHz. The spectral index above the peak is shallow ($\beta\approx-0.2$) with a flattening in the mm-band ($\gtrsim100$~GHz), as confirmed by extrapolations (open circles) of broken power law model fits (lines) to the observations at $\lesssim50$~GHz. 
    Right: Evolution of peak flux and peak frequency from broken power law fits to the radio SED at $\lesssim50$~GHz (Table~\ref{tab:radiobplfits})
    }
    \label{fig:radiofitev}
\end{figure*}

Fitting the radio SEDs at $\lesssim50$~GHz with fiducial, broken power-law models of the form 
\begin{equation}
F_\nu = F_{\rm peak} \left[\frac12\left(\frac{\nu}{\nu_{\rm peak}}\right)^{-s\beta_1} +  
\frac12\left(\frac{\nu}{\nu_{\rm peak}}\right)^{-s\beta_2} 
\right]^{-1/s},
\end{equation}
separately in each epoch, we find evidence for a slowly decreasing peak flux ($f_{\nu,\rm peak}\propto t^{-0.70\pm0.02}$) and peak frequency ($\nu_{\rm peak}\propto t^{-0.49\pm0.02}$) with time (Figure~\ref{fig:radiofitev} and Table~\ref{tab:radiobplfits}). 
From these fits, we confirm that the spectral index above the peak is shallow, $\betaradio\approx-0.2$ at $\lesssim28$~days, steepening marginally to $\approx-0.4$ at $\approx52$~days, and not well constrained thereafter. An extrapolation of the cm-band spectrum to the mm-band under-predicts the 97.5~GHz flux density by $\approx20$--50\%. Furthermore, the mean spectral index between the ALMA (3mm) and SMA (1.3mm) bands over this period is fairly flat, $\beta_{\rm mm}=-0.02\pm0.13$. Thus, the mm-band emission cannot be easily subsumed into this additional radio component without either invoking additional high-frequency structure in the emission, or invoking additional sources of systematic uncertainties.

The spectral index between the cm-band peak in the first VLA epoch at $\approx3.46$~days at 2.5~GHz and the ALMA 3\,mm (97.5~GHz) observation is $\beta=-0.106\pm0.002$, which is shallower than the cm-band spectral index alone. We test whether phase decorrelation at $\gtrsim10$~GHz in the VLA observations could be responsible for a loss in observed flux density at higher frequencies by self-calibrating the highest-frequency K-band (25~GHz) observations in the first epoch at $\approx3.46$~days. This process significantly reduces the imaging residuals but only marginally increases the 25~GHz flux density by 4\%, whereas making this flux consistent with the $\beta\approx-0.1$ power law would instead require an increase of $\approx10\%$. We recognize that all radio observations are subject to a systematic flux density uncertainty of $\approx10\%$ from the flux calibration process, which is not incorporated into the analysis above. Thus, if the true cm-band flux were systematically higher by this amount at 25~GHz (but not at 2.5~GHz), then there remains a possibility that the cm-band and mm-band could yet be ascribable to the same additional component. 

\begin{deluxetable*}{cccccccccc}
\label{tab:radiobplfits}
\tablecaption{Radio SED fits and inferred outflow properties}
\tablehead{
\colhead{Time} & \colhead{$F_{\rm peak}$} & \colhead{$\nu_{\rm peak}$}      & \colhead{$\beta_1$} & \colhead{$\beta_2$} & \colhead{$s$} & \colhead{R} & \colhead{$\Gamma$} & \colhead{$E_{\rm min}$} & \colhead{$B_{\rm eq}$} \\
\colhead{(d)} & \colhead{(mJy)} & \colhead{(GHz)}  & & & & \colhead{($10^{18}$\,cm)} &  & \colhead{($10^{48}$\,erg)} & \colhead{(mG)}
}
\startdata
3.46 & $8.6\pm0.2$ & $1.67\pm0.02$ & $2.5^\dag$ & $-0.18\pm0.01$ & $1.6\pm0.1$ & $1.40\pm0.03$ & $9.4\pm0.1$ & $1.63\pm0.02$ & $12.43\pm0.02$\\
6.44 & $4.9\pm0.8$ & $1.12\pm0.30$ & $2.5^\dag$ & $-0.20^\dag$ & $0.91\pm0.27$ & $1.43\pm0.60$ & $6.9\pm1.4$ & $2.13\pm0.24$ & $11.9\pm0.05$\\
12.44 & $4.3\pm0.8$ & $1.19\pm0.25$ & $2.5^\dag$ & $-0.26\pm0.02$  & $1.8\pm1.2$ & $0.88\pm0.12$ & $3.9\pm0.7$ & $4.02\pm0.51$ & $18.2\pm0.6$\\
17.48 & $2.9\pm0.1$ & $0.77\pm0.06$ & $2.5^\dag$ & $-0.23\pm0.01$ & $1.4\pm0.3$ & $1.04\pm0.11$ & $3.6\pm0.2$ & $4.28\pm0.10$ & $12.2\pm0.1$\\
28.33 & $1.9\pm0.1$ & $0.56\pm0.03$ & $2.5^\dag$ & $-0.21\pm0.01$ & $1.3\pm0.2$ & $1.00\pm0.08$ & $2.8\pm0.1$ & $5.30\pm0.19$ & $10.99\pm0.09$\\
52.48 & $1.2\pm0.1$ & $0.57\pm0.12$ & $2.5^\dag$ & $-0.37\pm0.03$ & $0.6\pm0.1$ & $0.58\pm0.18$ & $1.5\pm0.2$ & $7.63\pm0.45$ & $18.3\pm0.6$\\
76.42 & $1.0\pm0.1$ & $0.45\pm0.07$ & $2.5^\dag$ & $-0.18\pm0.01$ & $1.0^\dag$ & $0.60\pm0.14$ & $1.3\pm0.2$ & $9.96\pm0.68$ & $15.9\pm0.4$ 
\enddata
\tablecomments{$^\dag$ Fixed. The last four columns list the equipartition radius, Lorentz factor, minimum energy (in the emitting region), and magnetic field, respectively, computed using the formalism of \cite{bdnp13} (see Section~\ref{text:equipartition}).}
\end{deluxetable*}

\begin{figure*}
    \centering    
    \begin{tabular}{cc}
        \includegraphics[width=\columnwidth]{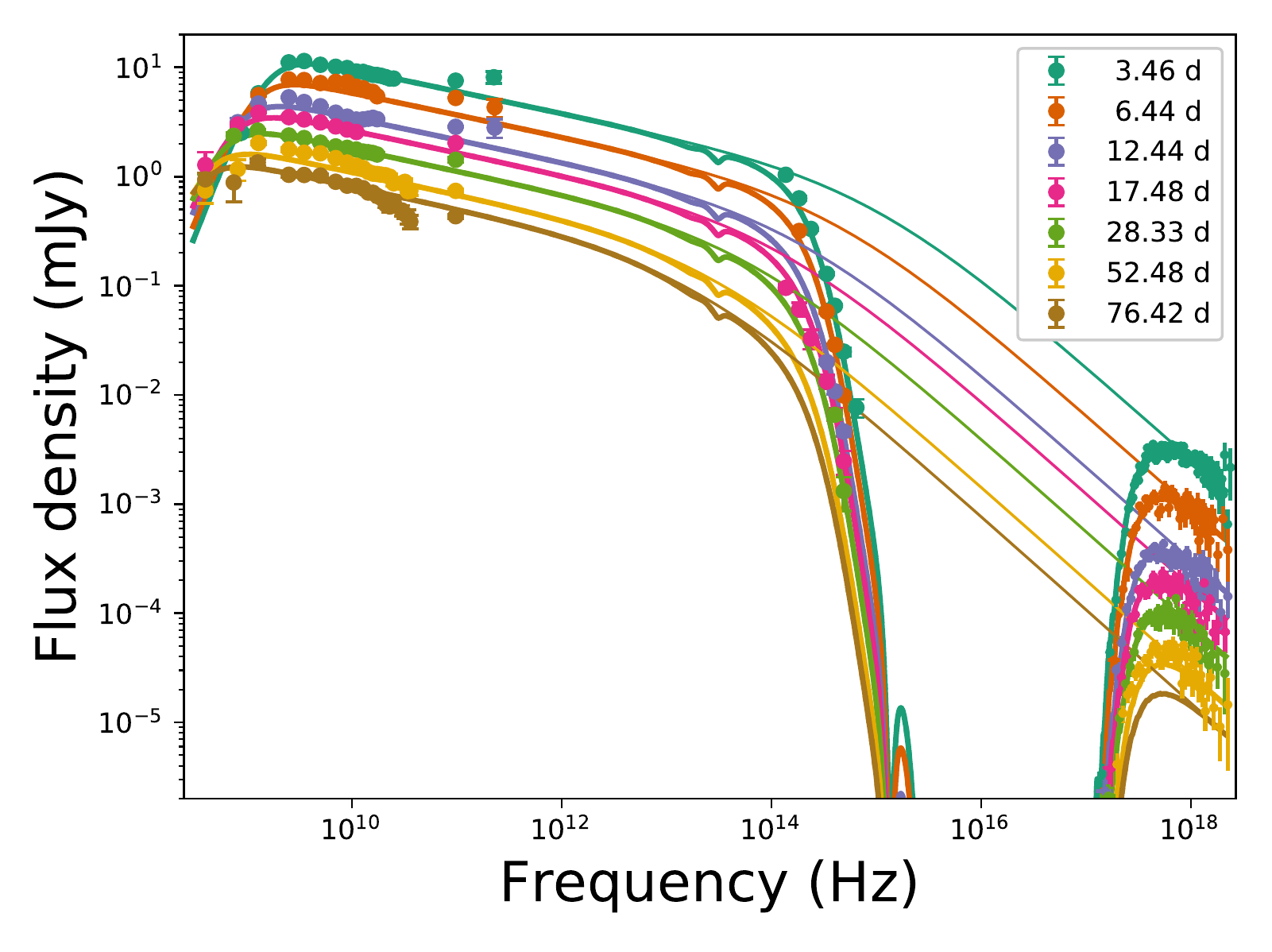} &
        \includegraphics[width=\columnwidth]{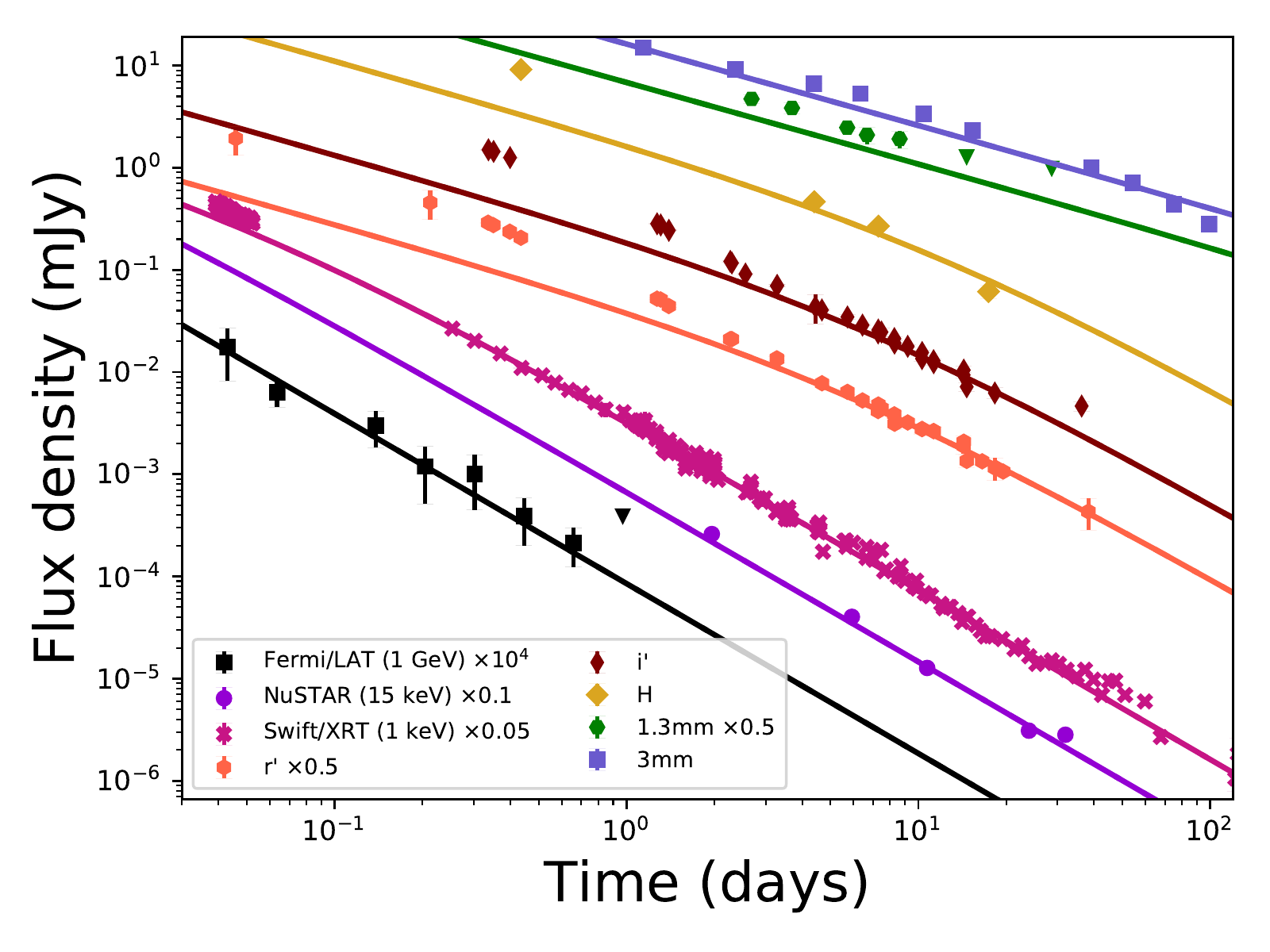}
    \end{tabular}
    \caption{Left: Radio to X-ray SEDs (same as Figure~\ref{fig:windfit}), together with a fiducial, single-component, broken power-law SED model (lines) where the three break frequencies evolve with time as $\nu_{\rm peak}\propto t^{-0.49}$, $\nu_{\rm b_1}\propto t^{-1.32}$, and $\nu_{\rm b_2}\propto t^0$, while the peak flux evolves as $F_{\nu,\rm peak}\propto t^{-0.7}$ (motivated by the observed radio evolution; Figure~\ref{fig:radiofitev}). The underlying model without the $\approx4.3$~mag of total extinction is shown as thin lines. Right: The corresponding light curves from the GeV to the mm band. The model under-predicts the optical observations at $\lesssim3$~days, the entire 1.3~mm light curve, and does not match any of the temporal breaks seen in the 3~mm light curve.
    }
    \label{fig:singlecomp}
\end{figure*}

We consider the possibility that the entire multi-frequency (radio to GeV) afterglow emission may in fact arise from a single emission component, but with non-standard evolution of break frequencies and fluxes. To test this, we anchor the peak of the SED (in $F_\nu$) as observed in the cm band, and evolve it in time according to the inferred peak frequency and peak flux evolution from the broken power-law fits to the radio SEDs (Table~\ref{tab:radiobplfits}). We assume spectral indices of $\beta_1=2.5$ and $\beta_2=-0.2$ below and above the peak, respectively. Upon extrapolating this spectrum to the optical, we find that an additional extinction of $A_{\rm V,host}\approx0.2$~mag is needed to match the optical flux, although this is still only possible at $\gtrsim3$~days. A spectral break is needed above the optical in order to not over-predict the X-ray flux. We find that an evolution of this break of $\nu_{\rm b_1}\approx 10^{15}\,{\rm Hz}\times(t/3.46\,{\rm d})^{-1.32}$ together with an assumed spectral index of $\beta_1\approx-0.86$ (to match the \nustar\ spectrum) above the break successfully reproduces the X-ray light curve at 1\,keV and 15\,keV. An additional break is then needed between the hard X-ray and GeV bands in order to not over-predict the GeV flux. Fixing the spectral index above this break to $\beta_2\approx-2$ (from the LAT spectrum), we find that the 1~GeV light curve can be reproduced by a fixed spectral break at $\nu_{\rm b_2}\approx20.7$~MeV. We plot this model in Figure~\ref{fig:singlecomp}. 

While this fiducial, single-component model adequately explains the observations in the cm-band, X-rays, and at 1\,GeV, it under-predicts all optical/NIR observations at $\lesssim3$~days (just like the physical wind-model; Figure~\ref{fig:fsfits-NuSTAR}) as well as the entire 1.3~mm (SMA) light curve. Furthermore, it does not have any of the temporal breaks that are evident in the 3~mm (97.5~GHz; ALMA \& NOEMA) light curve. It is possible that some of these limitations could be resolved by introducing additional, potentially moving breaks into the spectral shape; however, the introduction of these additional degrees of freedom would further reduce the predictive power of the model and make it even more challenging to interpret. Finally, there are no obvious explanations for this particular SED shape or evolution of break frequencies, and thus it is not straightforward to extract meaningful physical information from this model at this stage.  

\begin{figure*}
    \centering  
        \includegraphics[width=\columnwidth]{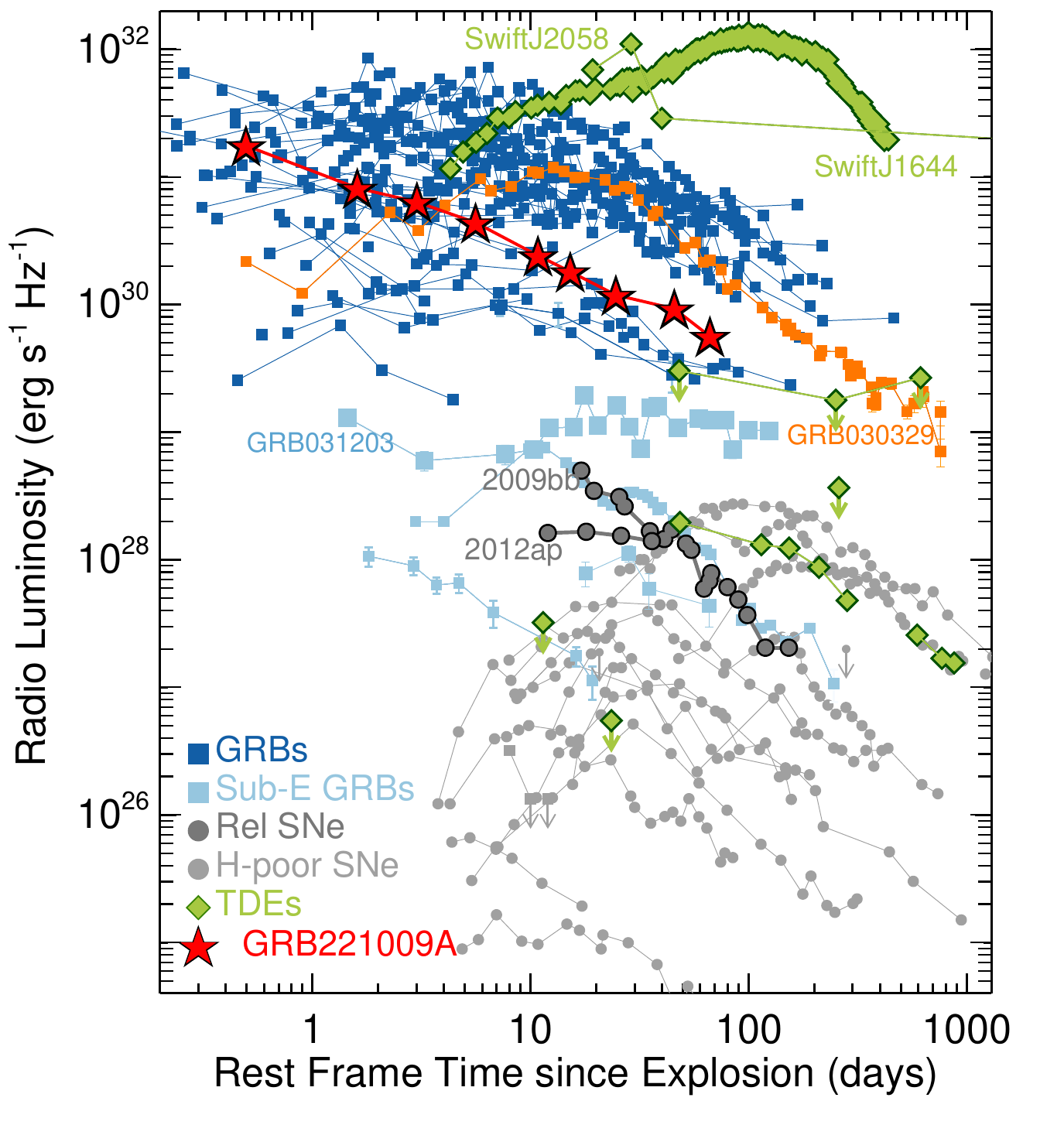}
         \includegraphics[width=\columnwidth]{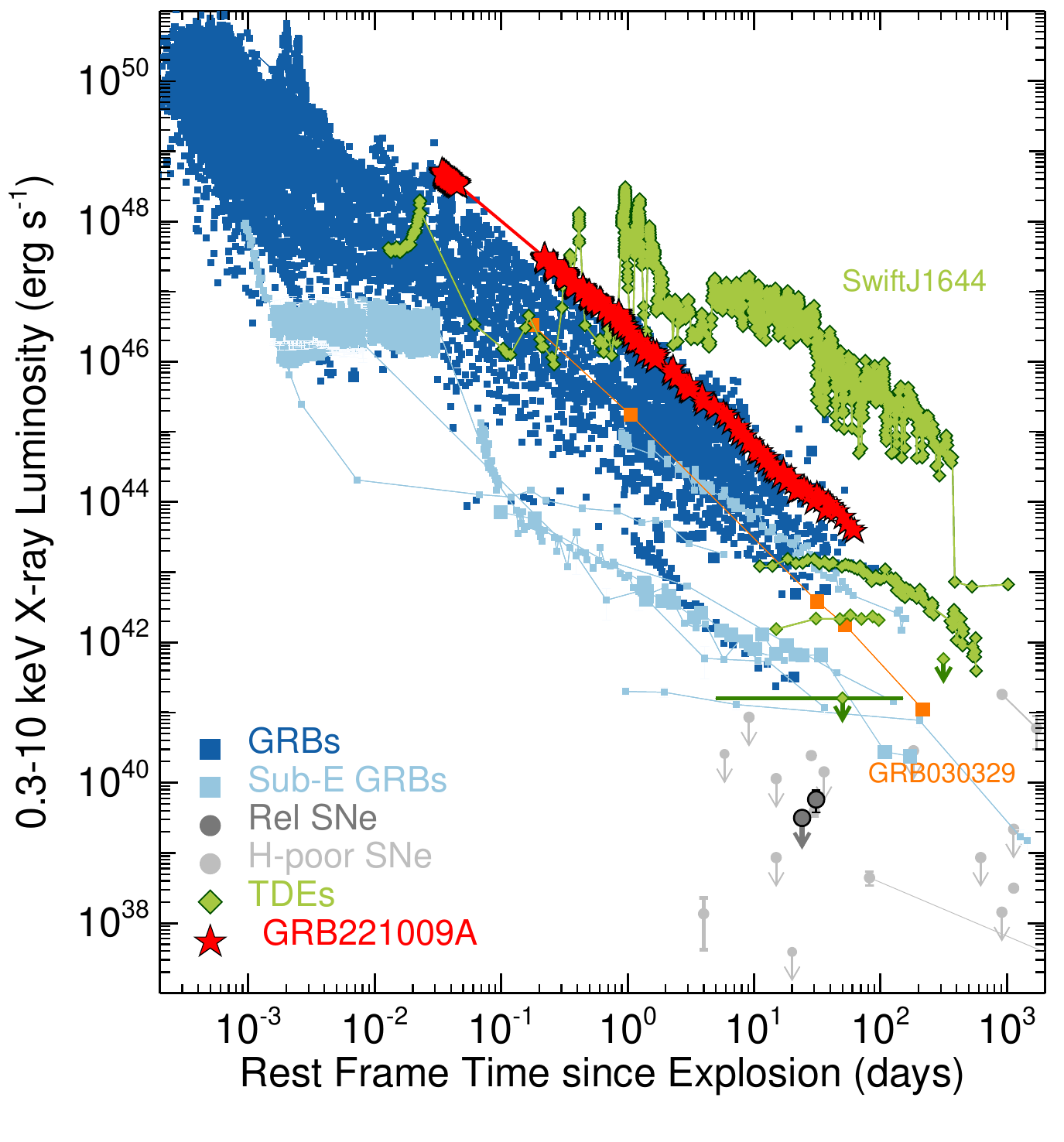}
    \caption{Left: Radio spectral luminosity at 8~GHz (rest frame) as a function of rest-frame time for GRBs (deep blue), sub-energetic GRBs (light blue), SNe with evidence for relativistic ejecta (dark gray) and hydrogen-poor supernovae (light grey). The radio afterglow of \grb\ is more luminous than any known radio supernova and comparable in luminosity to the afterglows of typical long GRBs. Right: The same plot showing X-ray luminosity in the 0.3-10 keV band. In contrast to the radio, \grb is one of the most X-ray luminous GRBs for much of its evolution. Adapted from \cite{mms+14}.}
    \label{fig:cmlc}
\end{figure*}

In the scenario that the excess cm-band (and possibly mm-band) emission arises from a separate population of radiating electrons, potential physical sources for such a component might be: 
(i) a radio supernova; 
(ii) a relativistic Maxwellian population of electrons (i.e, the ``non-accelerated electrons'' or ``thermal'' electrons \citep{ew05,rl17,wbi+18,mq21}; 
(iii) the reverse shock (RS); and
(iv) a two-component jet (i.e., two FS-like regions of different geometries), possibly with energy injection or time-varying microphysical parameters. The radio component is $\gtrsim100$ times more luminous than the brightest known radio SNe (Figure~\ref{fig:cmlc}) and peaks significantly earlier ($\sim$few days versus $\sim100$ days post-burst); we therefore do not consider emission associated with a possible SN to be a viable explanation for the radio excess in \grb.  
The emission spectrum from thermal electrons is expected to be broad and to cut off steeply above the peak \citep{rl17}. A detailed test against thermal electron models was performed by \cite{lves+19} for the low-frequency radio excess in GRB~181201A, and they confirmed that the radio SEDs in that case were narrower than thermal electron models would predict. The SEDs observed in this case are similarly sharply peaked, and thus also unlikely to match our current framework of synchrotron radiation from a Maxwellian population of electrons. 

In the case of GRB~181201A, the excess radio emission was ultimately ascribed to the RS, albeit with non-standard parameters. However, the simple power-law fits performed above demonstrate that the temporal evolution of \grb's radio component is too slow to be ascribed to a RS, {even with extreme parameters}. To see this, the evolution of the spectral peak\footnote{{The peak cannot be $\numr$ or $\nucr$, as these evolve as $t^{-1.5}$, which is too fast to match the data.}} ($\nuar$) of a Newtonian \footnote{The emission can decay slower in a Newtonian RS compared to the relativistic RS case.} RS in the regime $\numr<\nuar$ is $\frac{\partial\log{\nuar}}{\partial\log t}=-\frac{(15g+24)p+32g+40}{(14g+7)p+56g+28}$, where $\Gamma\propto R^{-g}$ is the evolution of the Lorentz factor of the post-shock ejecta with radius \citep{ks00}. We expect $0.5<g<1.5$ for a wind environment, but even if we set this aside and consider arbitrarily large values of $g$ (which results in a slower evolution), the expected temporal evolution of the peak frequency asymptotes to 
$-\frac{15p+32}{14p+56}\approx-0.7$ for the value of $p\approx1.4$ that is required to match the spectral index above the peak. Similarly, the evolution of the peak flux\footnote{The spectral index in this expression is not $5/2$ due to the definition of $\fnumr$ used here, which corresponds to the non-self-absorbed flux density at \numr.}, $\fnuar=\fnumr\left(\frac{\nuar}{\numr}\right)^{(1-p)/2}$. This gives $\frac{\partial\log{\fnuar}}{\partial\log t}=
-\frac{g(15p^2+42p-2)+4(6p^2+14p-5)}{7(2g+1)(p+4)}\approx-0.94$ for large $g$ and $p\approx1.4$. Both these result in a faster fading SED than observed in the cm band for this burst. For completeness, we present sample plots of the RS model {in the alternative regime of $\nuar<\numr<\nucr$} in Appendix~\ref{appendix:RS}. 

Finally, the radio flux cannot be easily ascribed to FS-like emission from a single power-law distribution of electrons either (e.g., in a two-component jet model). This is similar to the scenario explored by \citealt{smoy22} (although as mentioned earlier, their specific model is ruled out by the radio evolution at $\gtrsim5$ d). For $p\approx1.4$ for such a model we find that, using the relations of \cite{dc01} for the regime $1<p<2$, we would expect a peak frequency evolution in this case of $\nua\propto t^{0.34}$ for a spherical evolution, and $\nua\propto t^{0.93}$ for a jet (i.e., post jet-break), together with peak flux evolution of $F_{\rm peak}\propto t^{-1.2}$ and $F_{\rm peak}\propto t^{-1.9}$, respectively, none of which match the observations. On the other hand, while energy injection into the shock producing this emission may arrest the decay of this component, energy injection by itself would not naturally also produce a shallow spectrum above the peak. A combination of both $p<2$ and energy injection might be a feasible match; however, such a model appears somewhat contrived. Whereas time-varying microphysical parameters could result in a non-standard evolution of the spectral peak (usually discussed in terms of $\numax$, although the peak is at $\nua>\numax$ here), this would not explain the shallow spectrum above the peak either \citep{pk04,vdhpdb+14,mrk+21}. 

In summary, to our knowledge, the cm and mm emission in GRB 221009A do not correspond to the evolution of any standard emission component, including standard prescriptions for forward and reverse shock emission. This is the first time that such a component has been captured in such exquisite detail; however, previous radio observations of long GRBs have often been sparse, leaving the possibility open that similar emission may be common in long GRBs. We discuss the potential prevalence of such an emission component in radio afterglows of GRBs in Section~\ref{text:radiocomparison}.

\section{Discussion}
\subsection{Comparison to other GRBs}
\label{text:radiocomparison}

\begin{figure}
    \centering    
        \includegraphics[width=\columnwidth]{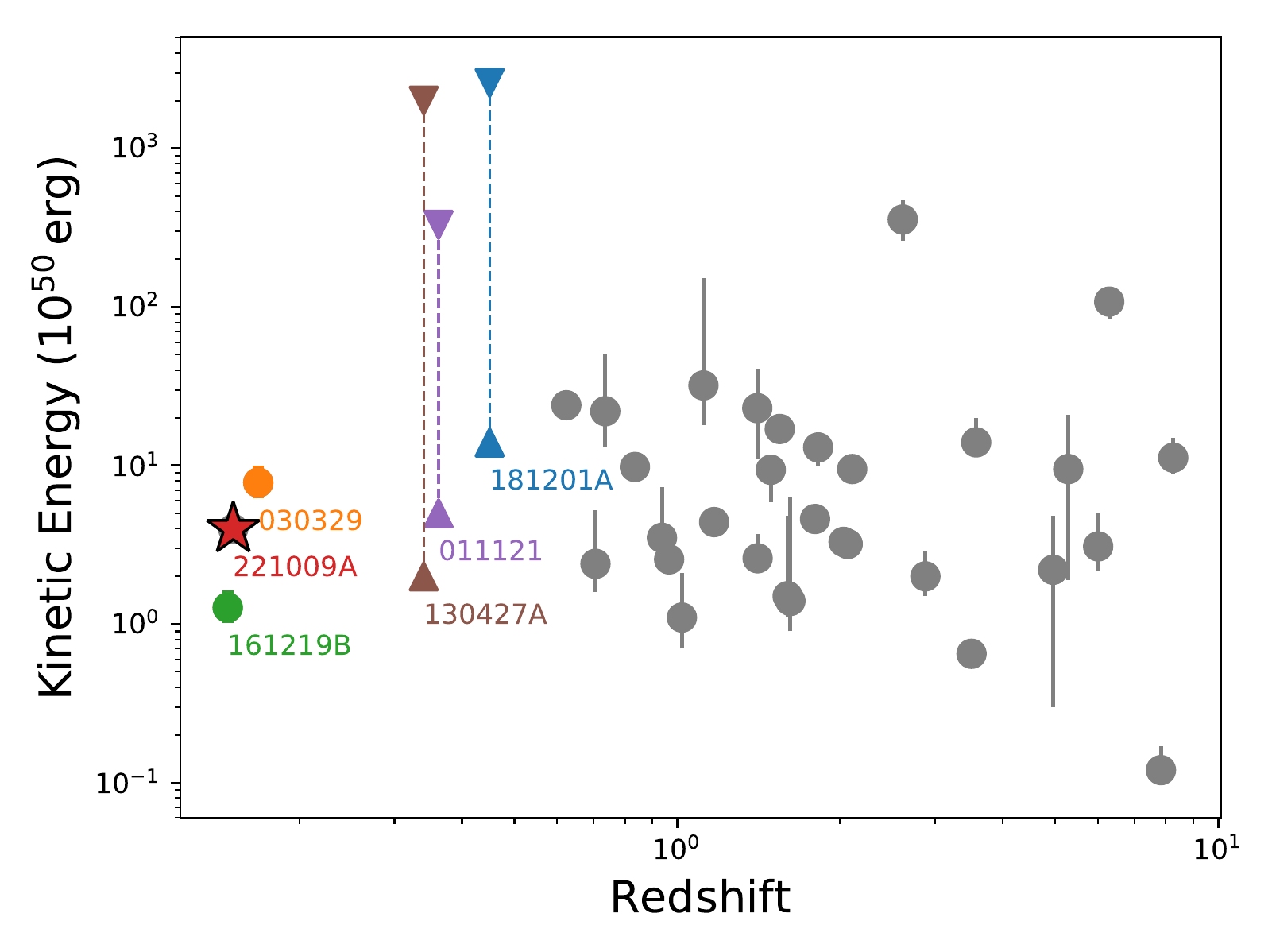} 
    \caption{Beaming-corrected kinetic energies of GRBs with multi-wavelength datasets and known jet opening angles, as a function of redshift, {together with the value of \EK\ for \grb\ derived from our best-fit FS model}.
    {The extraordinary brightness of \grb\ at high energies may arise from a combination of its proximity and a narrow beaming angle, rather than a high intrinsic \EK.}
    }
    \label{fig:EK_z}
\end{figure}

We now consider \grb\ in the context of the broader GRB population. In the X-ray and $\gamma$-ray bands, \grb\ is one of the brightest GRBs ever observed \citep{gcn32636,gcn32650,gcn32651,gcn32668,gcn32694,gcn32756,ndlo+23}. While this is due in part to its proximity, \grb\ is also intrinsically among the most luminous known bursts at these wavelengths (Figure \ref{fig:cmlc}). 
Our FS modeling of \grb\ suggests that this superlative luminosity is {likely} not due to an unusually powerful jet, but rather to fortuitous geometric alignment: the beaming-corrected jet kinetic energy is average for long GRBs, $E_K=50.60^{+0.09}_{-0.07}\,\rm{erg}$, but the jet's small opening angle, $\theta_{jet}=1.64^{+0.28}_{-0.20}$, places it among the most narrowly collimated jets in the GRB population \citep{lab+18}. {We note that the jet break at $\approx130$~days in our model is driven by the mm-band light curve, and, if a more accurate multi-wavelength model can be found that also successfully incorporates the radio and mm observations, then this interpretation would likely need to be revisited. Under this FS model, }\grb's kinetic energy is similar to that of other low-redshift GRBs; in particular, it is intermediate between GRB 161219B and GRB 030329 (Figure \ref{fig:EK_z}). Given the high $\gamma$-ray energy, this implies a relatively high prompt efficiency of $\eta_{\gamma}\approx86\%$, {independent of the beaming correction}. 

\begin{figure}
    \centering    
        \includegraphics[width=\columnwidth]{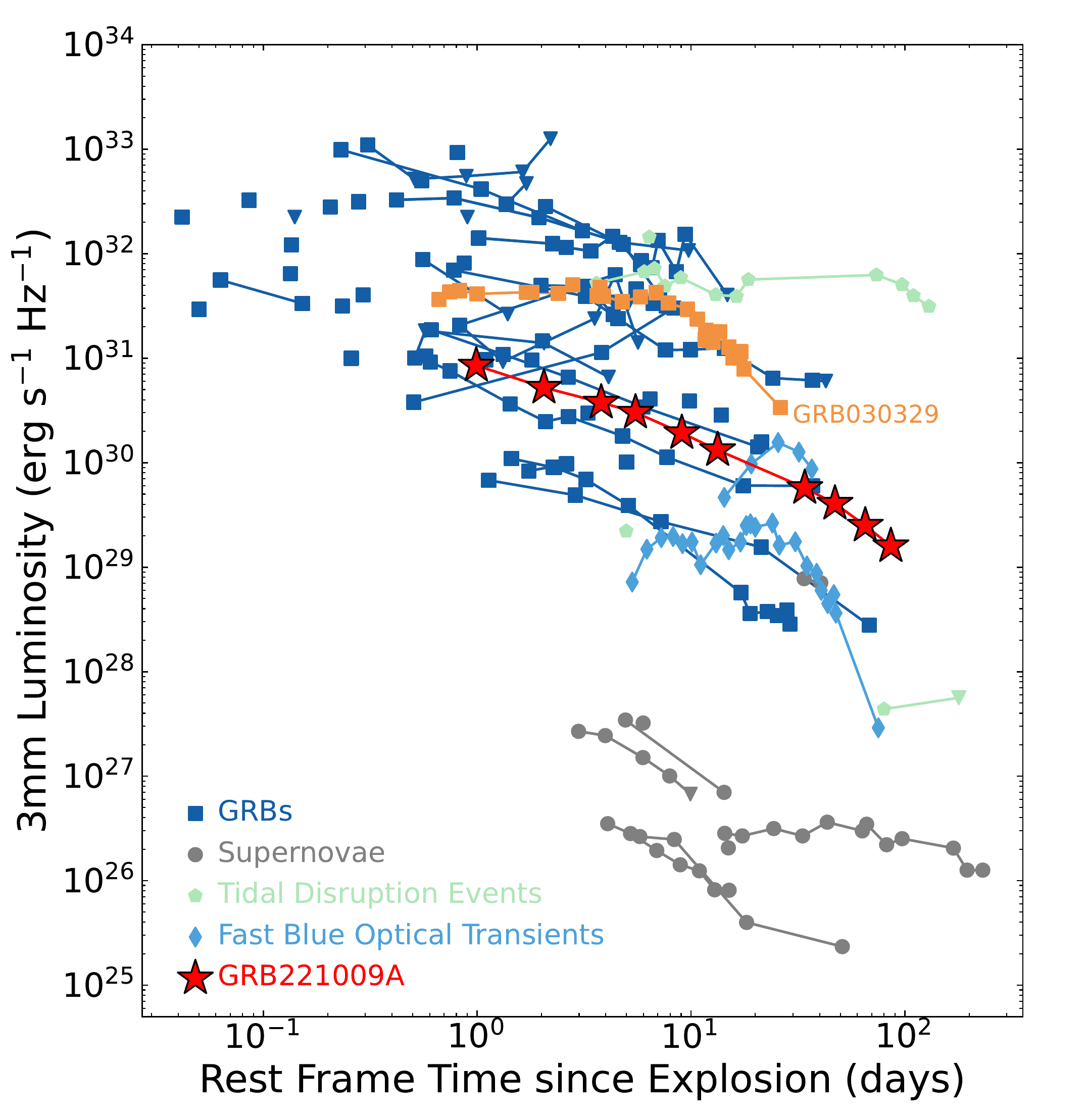} 
    \caption{Spectral luminosity of \grb\ as a function of rest-frame time in the 3mm band (red stars), compared to a population of long-duration GRBs (deep blue), supernovae (grey), tidal disruption events (light green), and fast blue optical transients (light blue). No K-corrections for variations in intrinsic spectral shape have been applied. The mm-band emission from \grb\ is a factor of $\approx5$--10 less luminous than GRB~030329 (orange), but of comparable luminosity to that of other GRB afterglows. Adapted from \cite{ebm+22}.}
    \label{fig:mmlc}
\end{figure}

\grb\ is also superlative in the quality and coverage of radio and mm data obtained. Despite \grb's extreme brightness at high energies, its cm and mm emission is merely average for the GRB population (Figures \ref{fig:cmlc} and \ref{fig:mmlc}). It is this combination of extreme X-ray (and optical) luminosity and mm mediocrity that makes fitting a FS model to the full dataset so challenging: the model overpredicts the mm emission at late times unless the mm band remains below the peak of the SED, forcing $\nu_m$ to remain between the mm and optical bands throughout the duration of our observations and requiring a jet break at $\approx130$~days. {Since the inferred narrow collimation angle for this burst is largely constrained by the mm-band light curve, most of which our model cannot explain, a more complete description of the radio emission is also required to derive a more accurate jet opening angle.}

Radio observations can also provide unique insight into the physical composition of the jet itself via the detection of RS emission \citep{lbz+13,pcc+14,lab+16,alb+17,lves+19,lag+19}. However, due to the frequent paucity of radio data, deviations of individual data points from this basic picture are often ignored, or attributed to other effects such as interstellar scintillation (e.g., \citealt{alb+17,alb+19,bhvdh+19}). As discussed in Section \ref{text:radiocomponents} and the Appendix, the exquisite temporal and frequency sampling of the radio dataset collected for \grb\ rule out a standard RS+FS picture, as well as other commonly considered model extensions (a two-component jet/jet+cocoon, emission from thermal electrons), with high confidence. We  attempt to determine if similar behavior could have been previously overlooked in other well-studied radio GRB afterglows in the literature. 

Recently \cite{kf21} systematically considered multi-wavelength observations of 21 well-studied long GRBs. They showed that while apparent deviations from a simple FS or FS+RS model in the radio within a single event may not appear statistically significant, when the population is considered as a whole, about half of the sample is difficult to explain with a standard afterglow model. \cite{ldf+23} obtained similar results based on a radio-only analysis of a slightly larger sample. In particular, several GRBs in \cite{kf21}'s sample (GRB~141121A, GRB~160625B, and GRB~171010A) exhibit shallow radio SEDs at late times, similar to \grb. Notably, GRB~141121A and GRB~160625B have early radio emission consistent with a standard RS model, but the data are brighter than the models at late times, similar to the issue we faced in attempting to model \grb's radio emission with a RS. These three GRBs are among the few published events with multi-frequency radio coverage extending to such late times, suggesting that broad, slowly-evolving radio components like that seen in \grb\ may be more common than previously realized, and may occur in GRBs both with and without distinct RS emission at early times.

While the number of GRBs with well-sampled mm light curves remains small, several of them also exhibit puzzling behavior relative to other wavebands. For example, GRB~161219B and GRB~181201A's mm light curves were both modeled as the sum of RS and FS emission, but fits to the mm light curves in isolation preferred a single power law decline with a temporal decay rate intermediate between the optical and cm bands, rather than a broken power law \citep{lab+18,lag+19}. Similarly, \grb's 3mm light curve evolves as $\sim t^{-1.1}$ at late times -- steeper than the low-frequency radio light curves ($\sim t^{-0.7}$ at 1.4 GHz), but shallower than the late-time optical and X-ray light curves. This together with the cm-to-mm SED shape further suggests that the mm emission in \grb\ may be dominated by yet another separate component, in addition to the FS that dominates the optical and X-rays and the slowly evolving component that dominates the cm bands. However, standard synchrotron theory struggles to produce emission that is narrowly peaked enough to dominate the mm while remaining sub-dominant at all other frequencies throughout the duration of our observations.

\subsection{Equipartition analysis of the radio component}\label{text:equipartition}

Assuming that the emission dominating the cm bands arises from synchrotron radiation from a shock powered by an outflow, we use energy equipartition arguments to derive estimates of the emission radius, minimum energy, equipartition magnetic field, and bulk Lorentz factor of the emitting region. We employ the formalism of \cite{bdnp13}, assuming area and volume-filling factors of unity, and present our results in Table~\ref{tab:radiobplfits}. The emission radius and equipartition magnetic field are roughly constant at $R\approx10^{18}$~cm and $B\approx13$~mG, respectively, whereas the bulk Lorentz factor decreases from $\Gamma\approx9.4$ to $\Gamma\approx1.3$ between $3.46$--76.42 days. This is lower than the corresponding bulk Lorentz factor of the FS (which decreases from $\Gamma\approx73$ to $\Gamma\approx34$ over the same period), possibly suggesting some structure in the ejecta. The minimum energy increases\footnote{In the equipartion framework, this apparent increase in $E_{\rm min}$ is due to deceleration resulting in a greater volume of the emitting plasma becoming visible, with values computed at later epochs providing a stronger constraint on the true minimum energy.} from $E_{\rm in}\approx1.6\times10^{48}$~erg to $E_{\rm min}\approx1\times10^{49}$~erg over this period. 

The apparent decelerating nature of this component yields a lower limit on the Lorentz factor of $\Gamma_0\approx9.4$, which corresponds to an upper limit\footnote{The relationship $E\approx\Gamma M_{\rm ej}c^2$ holds only if the system is still ballistic. If it has been decelerated by the environment (as the stationary radius would indicate), then the Lorentz factor should no longer be connected to the ejecta mass, but only to the energy and density. Since the Lorentz factor decreases with time, and because we infer an increasing energy and decreasing $\Gamma$, the first epoch yields the tightest constraint and an upper limit on $M_{\rm ej}$.} on the ejecta mass of $M_{\rm ej}\approx E_{\rm min}/(\Gamma_0 c^2)\approx10^{-7}~M_{\odot}$ for $E_{\rm min}\approx1.6\times10^{48}$~erg as derived from the radio SED at 3.46~days, or  $M_{\rm ej}\lesssim6\times10^{-7}~M_{\odot}$ as derived from the radio SED at 76.42~days. This is similar in magnitude to the inferred ejecta mass for a typical long-duration GRB with beaming-corrected kinetic energy $\EK\approx5\times10^{50}$~erg \citep{lbt+14}, assuming a typical initial Lorentz factor of $\Gamma_0\approx100$. This indicates that the outflow producing the radio emission in \grb\ shares characteristics similar to those of standard long-duration GRBs. However, the stagnant emission radius inferred for this component stands in stark contrast to the expanding outflow expected from standard theoretical models of a relativistic GRB jet propagating in a smooth environment. Finally, the inferred value of the equipartition magnetic field is higher than that inferred for the post-shock B-field in the FS ($B\approx1~{\rm mG}(t/1~{\rm day})^{-3/4}$), but lower than that inferred for supernova shocks in Type Ib/c supernovae \citep{cf06}.

\subsection{Caveats from modeling the FS emission}
\label{text:fscaveats}
We now discuss the mismatch between the FS wind model and the data at multiple wavelengths. The FS model under-predicts the optical emission at $\lesssim1$~day (Figure~\ref{fig:fsfits-NuSTAR}). Similar excess emission in the past has been attributed to RS emission, although the RS in this case would not match the radio observations (Appendix~\ref{appendix:RS}). A change in density structure from steep to shallow remains plausible, although the density profile is already steep ($k=2$) and the density itself already very low ($A_*\approx2\times10^{-3}$). Furthermore, such a transition would also affect the X-rays, which are in the same spectral regime in this model; however, no such transition is apparent in the light curves. Any additional component invoked to explain these optical observations would either need to match the radio SED, or at least, not over-predict these. Given the lack of such models, the observed optical excess is puzzling. 

The FS model also under-predicts the optical emission at $\gtrsim30$~days. Excess emission above this model is also apparent in the \swift/XRT light curve at 1~keV in the form of a bump at 30--60~days, and in the final \nustar\ epoch relative to the single power-law fit to the 15~keV light curve (Figure~\ref{fig:lcbplfits}). In fact, a similar excess over the broken power-law fits is also apparent in the radio light curves on a very similar timescale (Figure~\ref{fig:lcbplfits}), possibly indicative that the radio emission does arise from the same emission region as the X-rays and optical, at least in this narrow time interval. The luminosity of this excess at 11~GHz, the optical $i'$-band, and at 1~keV is $\approx2\times10^{39}$~erg\,s$^{-1}$, $\approx5\times10^{41}$~erg\,s$^{-1}$, and $\approx10^{43}$~erg\,s$^{-1}$, respectively. The achromatic nature of this bump is suggestive of a hydrodynamic effect, such as the appearance of a counterjet, a second emission component (e.g., cocoon or SN emission), a density enhancement (possibly including an encounter with the wind termination shock), or energy injection. We expect $\Gamma\approx40$ at $\approx60$~days, so the counterjet possibility appears unlikely. The luminosity of this component is greater than any known SN at each of these wavelengths (Figure~\ref{fig:cmlc}), making such an origin also unlikely. Previous work on the impact of density enhancements on optical light curves of GRB afterglow suggests that such effects lead to minimal deviations in the light curves owing to the highly relativistic nature of the jet, even if the enhancement is extreme \citep{ng07,ub07,vemwk09,velm+10,gvem13,gwl+14}. Whereas energy injection could conceivably create such a signature, it would need to be coupled with a jet-break in order for the light curves to not asymptote to a higher flux level. In their study of energy injection in GRBs, \cite{lbm+15} find that in 3 out of 4 cases studied, the jet breaks were within a factor of $\approx2$ from the end of the period of energy injection. If the observed bumps in the light curve for \grb\ are related to energy injection, this would continue such a trend. However, the inability of the FS model to directly connect with the radio observations makes further investigation of this possibility challenging. 

The best-fit wind model over-predicts the observed NIR ($JHK$-band) flux by $\approx15$--30\%, with the offset worsening with time. On the other hand, no such offset is apparent in the optical fits. We find that this is due to an apparent change in the spectral slope between the NIR and optical bands from $\betaniropt=-0.76\pm0.04$ at $\approx4.4$~days (Section~\ref{text:fs}) to $\betaniropt=-0.52\pm0.04$ at $\approx17.5$~days. Such a hardening of the NIR-to-optical spectrum is unexpected and cannot be understood in the standard synchrotron framework in our preferred spectral regime of $\numax\lesssim\nuopt<\nux<\nuc$. Whereas such a color evolution from red to blue is similar to that expected from the emergence of an underlying SN, inclusion of such a component would over-predict the optical observations. Suppressing the afterglow contribution to both the optical and NIR at this time would require either a steepening of the electron index, $p$ with time (which would then affect the X-ray fit) or a faster evolution of $\numax$, which would worsen the tension in the mm-band at $\approx100$~days, making this interpretation untenable. 
It is also possible that this apparent change in $\betaniropt$ instead arises from systematic calibration errors in the data. Further investigation of this anomaly requires better light curve coverage of the NIR afterglow and is beyond the scope of this work. 

The model fit appears slightly worse in the LAT 1~GeV band. This is partially due to a slightly shallower decay index -- the model predicts $\alphaLAT=(2-3p)/4\approx-1.40$, compared to the observed value of $\alphaLAT\approx-1.47$. It is also in part due to the higher flux in the model (by $\approx50\%$) compared to the data. The flux offset could be addressed by a slightly lower value of $\nuc$, or, alternatively, both of these could be remedied by a slightly larger value of $p$. Since $\nuc$ depends on all four physical parameters, it is non-trivial to discern the origin of additional constraints on its value imposed by the rest of the data. The challenge with a higher value of $p$, on the other hand, lies in fitting the optical-to-X-ray spectral index; increasing $p$ while keeping the model flux in the X-rays fixed would result in the model over-predicting the NIR even more. Thus the current value of $p$ is a compromise between strongly over-predicting one or the other of the NIR or GeV emission. Finally, we note that the LAT spectrum shows some evidence for steepening in the last two bins at $\gtrsim31$~ks ($\gtrsim0.4$~days). This could be due to the lower photon flux reducing the fluence of the highest energy photons in this bin. However, if this reduction is real, it could also indicate the movement of the maximum synchrotron frequency, $\nu_{\rm max}$ into the LAT band. We defer the discussion of this potential effect to future work. 

Finally, we speculate on a possible unifying underlying cause for some of these issues. One possible way to resolve the tension between our observations and our analytic model might lie in constructing a more realistic distribution of relativistic electron energies. Standard synchrotron theory assumes that the population of radiating electrons are accelerated into a simple power-law distribution of energies with an abrupt low-energy cutoff at $\gamma_{\rm min}$, resulting in a spectral break ($\nu_m$). Our current model for \grb\ requires $\nu_m$ to remain between the mm and the optical throughout our observations; we never observe a break in any light curve attributable to the transition of $\nu_m$ through that frequency band, although we infer its presence in the optical at $\lesssim2$~days by dint of the constraint on the flux of the $\nu^{1/3}$ segment imposed by the mm-band data at $\approx100$~days. Changing the low-energy end of the relativistic electron distribution would change the SED below $F_{\rm peak}$, exactly the regime where our current model struggles to reproduce our observations. We defer further exploration of this and other possibilities to future work.

\subsection{Synchrotron Self-Compton predictions}
Multiple experiments reported the detection of very high energy photons from \grb\ \citep{gcn32677,ATel15669}, making \grb\ the newest member of the very small class of GRBs with detected VHE emission (GRBs~180720B, 190114C, 190829A, and 201216C; \citealt{mcaa+19a,aaa+19,gcn29075,hcaa+21}). \grb\ also exhibited the highest-energy photon yet associated with any GRB (18~TeV; \citealt{gcn32677}). Armed with a predictive model, however imperfect, we consider whether these VHE photons could arise from FS emission. \cite{gcn32677} report 5000~VHE photons in the span of $\approx2000$~s following the \fermi/GBM trigger, corresponding to a flux of $\approx3\times10^{-10}$\,erg\,s$^{-1}$\,cm$^{-2}$ at 1~TeV (computed assuming a LHAASO collecting area of 1\,km$^2$ and a mean photon energy of $\approx1$~TeV, ignoring spectral corrections). At a time of $\approx500$~s after the burst, we compute both the synchrotron spectrum from our FS model, and the corresponding synchrotron self-Compton (SSC) emission expected, the latter by integrating the synchrotron spectrum over an electron distribution back-calculated from the locations of the break frequencies at this time. We find that a synchrotron flux and SSC flux, at 1~TeV, of $\approx1.2\times10^{-8}$\,erg\,s$^{-1}$\,cm$^{-2}$ and $\approx3\times10^{-11}$\,erg\,s$^{-1}$\,cm$^{-2}$, respectively. Thus, {in the (presumed) absence of a high-energy cutoff in the electron spectrum}, the synchrotron spectrum will dominate over SSC at 1~TeV, and hence the intrinsic spectrum is expected to be $\approx-p/2\approx-1.1$ in $F_\nu$. The synchrotron flux is a factor of $\approx40$ higher than the rough observed flux computed above; absorption due to $\gamma$-$\gamma$ pair production against the Extragalactic Background Light (EBL) is expected to attenuate the observed spectrum (although, see also \citealt{smsr23}), and this deficit is of the same order of magnitude as, but smaller than, that inferred for GRB~190114C \citep{mcaa+19a}. Thus, it is conceivable that the VHE emission for this GRB was produced by the FS, {although a full analysis requires the VHE data}. We conclude with two caveats. First, this is only an order-of-magnitude estimate, and have not considered, for instance, a high-energy cutoff in the electron spectrum in this calculation {(which could lead to SSC dominating over synchrotron emission in the VHE range)}. Second, our FS model has several shortcomings, and the true FS model flux at the time of the VHE detection remains somewhat uncertain. 

\section{Conclusions}
We have presented multi-wavelength observations of the superlative \grb, spanning fifteen orders of magnitude in frequency and four in time. We find that \grb's NIR, optical, X-ray, and $\gamma$-ray emission can be well-modeled as a synchrotron FS from a highly collimated relativistic jet interacting with a low-density wind-like medium. \grb's high brightness as observed from Earth can be attributed to a combination of its relative proximity, its large intrinsic luminosity, and{, potentially, a} high degree of jet collimation combined with an on-axis orientation. 

\textit{\grb\ strongly demonstrates the need for additional theoretical work to fully understand the ultra-relativistic jets seen in long GRBs.} While a simple FS model is broadly consistent with a large fraction of our data, the radio and mm emission in particular are difficult to explain within the scope of standard synchrotron theory. We consider two possibilities: (i) that the radio emission is due to an additional synchrotron emission component (with a possible second additional component required to explain the mm emission) or (ii) that our basic analytic models of relativistic synchrotron emission need to be modified in some fundamental way. We find that the temporal evolution and spectral shape of the cm emission are inconsistent with standard analytic models for FS or RS emission propagating in constant density or wind-like media. However, the peak frequency and peak flux density of this component evolve simply with time (as $\nu_{\rm peak}\propto t^{-0.5}$ and $F_{\rm peak}\propto t^{-0.7}$ respectively), perhaps suggesting that an analytic description of this component might be possible if a non-standard assumption is made (e.g., evolving microphysical parameters). While fully exploring extensions to the standard synchrotron afterglow models is beyond the scope of this paper, we briefly speculate that a more realistic treatment of the low-energy end of the
relativistic electron distribution may solve some of the issues.

\grb's proximity means that it will remain detectable with a wide variety of radio facilities for years to come, providing a testbed for future theoretical work and an opportunity to further refine the synchrotron model applied in this paper. In addition, \grb's radio brightness and longevity will provide rich opportunities for additional science, including directly measuring the physical size of the afterglow with VLBI observations (previously only convincingly demonstrated for GRB~030329; {\citealt{tfbk04}}) and constraining the magnetic field structure of the jet with polarization observations {(such as for GRB~190114C; \citealt{lag+19})}. 

\section*{Acknowledgements}
We thank J.~ Racusin and E.~ Burns for contribution to the \nustar\ observations and for helpful comments. {We thank the anonymous referee for their rapid and constructive review of this work.}
T.E.\ is supported by NASA through the NASA Hubble Fellowship grant HST-HF2-51504.001-A awarded by the Space Telescope Science Institute, which is operated by the Association of Universities for Research in Astronomy, Inc., for NASA, under contract NAS5-26555. SB is supported by a Dutch Research Council (NWO) Veni Fellowship (VI.Veni.212.058).  
The work of RY is partially supported by JSPS KAKENHI (Grant No. JP22H01251).
R.B.D.\ acknowledges support from the National Science Foundation under grant 2107932. AG acknowledges the financial support from the Slovenian Research Agency (research core funding No. P1-0031, infrastructure program I0-0033, and project grant No. J1-8136, J1-2460). 
The TReX group at Berkeley is partially supported by NSF grants AST-2221789 and AST-2224255. 

GMRT observations for this study were obtained via project 43\_039 (PI: Laskar). We thank the staff of the GMRT that made these observations possible. GMRT is run by the National Centre for Radio Astrophysics of the Tata Institute of Fundamental Research. 
The MeerKAT telescope is operated by the South African Radio Astronomy Observatory, which is a facility of the National Research Foundation, an agency of the Department of Science and Innovation.
VLA and VLBA observations for this study were obtained via projects VLA/22B-062 
and VLBA/22B-305, respectively (PI: Laskar). 
The National Radio Astronomy Observatory is a facility of the National Science Foundation operated under cooperative agreement by Associated Universities, Inc.
The Australia Telescope Compact Array is part of the Australia Telescope National Facility (https://ror.org/05qajvd42) which is funded by the Australian Government for operation as a National Facility managed by CSIRO. We acknowledge the Gomeroi people as the Traditional Owners of the Observatory site.
This paper makes use of the following ALMA data: ADS/JAO.ALMA\#2022.1.01433.T. ALMA is a partnership of ESO (representing its member states), NSF (USA) and NINS (Japan), together with NRC (Canada), MOST and ASIAA (Taiwan), and KASI (Republic of Korea), in cooperation with the Republic of Chile. The Joint ALMA Observatory is operated by ESO, AUI/NRAO and NAOJ.
This work is based on observations carried out under project number S22BE with the IRAM NOEMA Interferometer. IRAM is supported by INSU/CNRS (France), MPG (Germany) and IGN (Spain). We thank Melanie Krips and the NOEMA staff for executing our observations, undertaking the reduction, and providing us with reduced data products. 
We thank the SMA staff for rapidly approving our ToO request and Mark Gurwell for reducing the data and providing us with the flux density measurements. 
The Submillimeter Array is a joint project between the Smithsonian Astrophysical Observatory and the Academia Sinica Institute of Astronomy and Astrophysics and is funded by the Smithsonian Institution and the Academia Sinica. We recognize that Maunakea is a culturally important site for the indigenous Hawaiian people; we are privileged to study the cosmos from its summit.

The Liverpool Telescope is operated on the island of La Palma by Liverpool John Moores University in the Spanish Observatorio del Roque de los Muchachos of the Instituto de Astrofisica de Canarias with financial support from the UK Science and Technology Facilities Council.
This work makes use of data supplied by the UK Swift Science Data Centre at the University of Leicester and of data obtained through the High Energy Astrophysics Science Archive Research Center On-line Service, provided by the NASA/Goddard Space Flight Center.
This work was supported under NASA contract No. NNG08FD60C, and made use of data from the NuSTAR mission, a project led by the California Institute of Technology, managed by the Jet Propulsion Laboratory, and funded by the National Aeronautics and Space Administration. This research has made use of the NuSTAR Data Analysis Software (NuSTARDAS) jointly developed by the ASI Science Data Center (ASDC, Italy) and the California Institute of Technology (USA).

\facilities{GMRT, MeerKAT, VLA, VLBA, ATCA, ALMA, NOEMA, SMA, Liverpool Telescope, \swift, \nustar, \fermi.}

\software{CASA \citep{mws+07}, FermiTools \citep{fss19}, XSPEC (v12.12.1; \citealt{arn96}), HEAsoft \citep{nhe14}, MIRIAD \citep{stw95}, AIPS \citep{gre03}, emcee \citep{fhlg13}, matplotlib \citep{hun07}.}

\appendix

\section{An ISM model}
\label{appendix:ISM}
\begin{figure*}
    \centering    
    \begin{tabular}{cc}
        \includegraphics[width=\columnwidth]{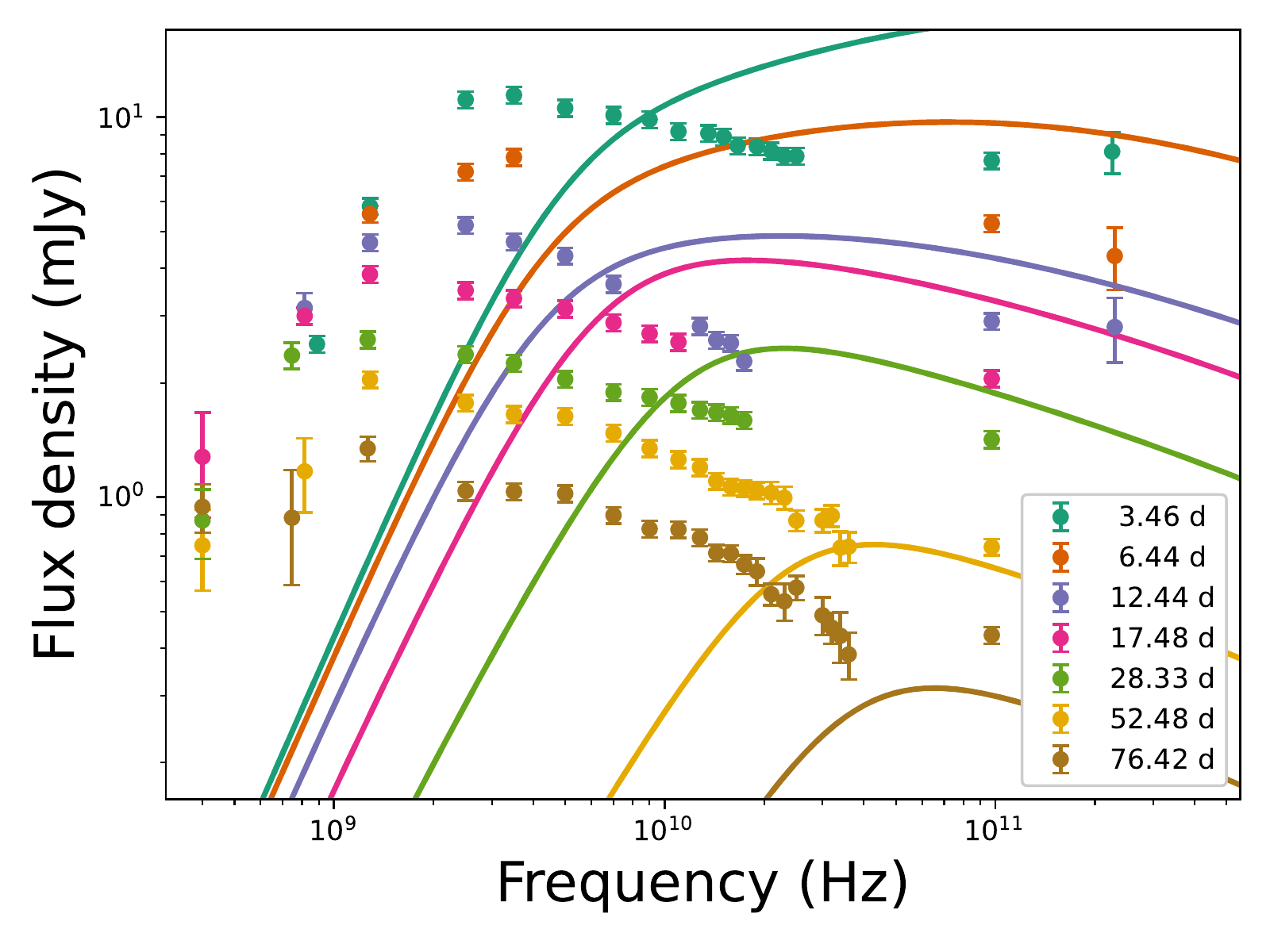} &
        \includegraphics[width=\columnwidth]{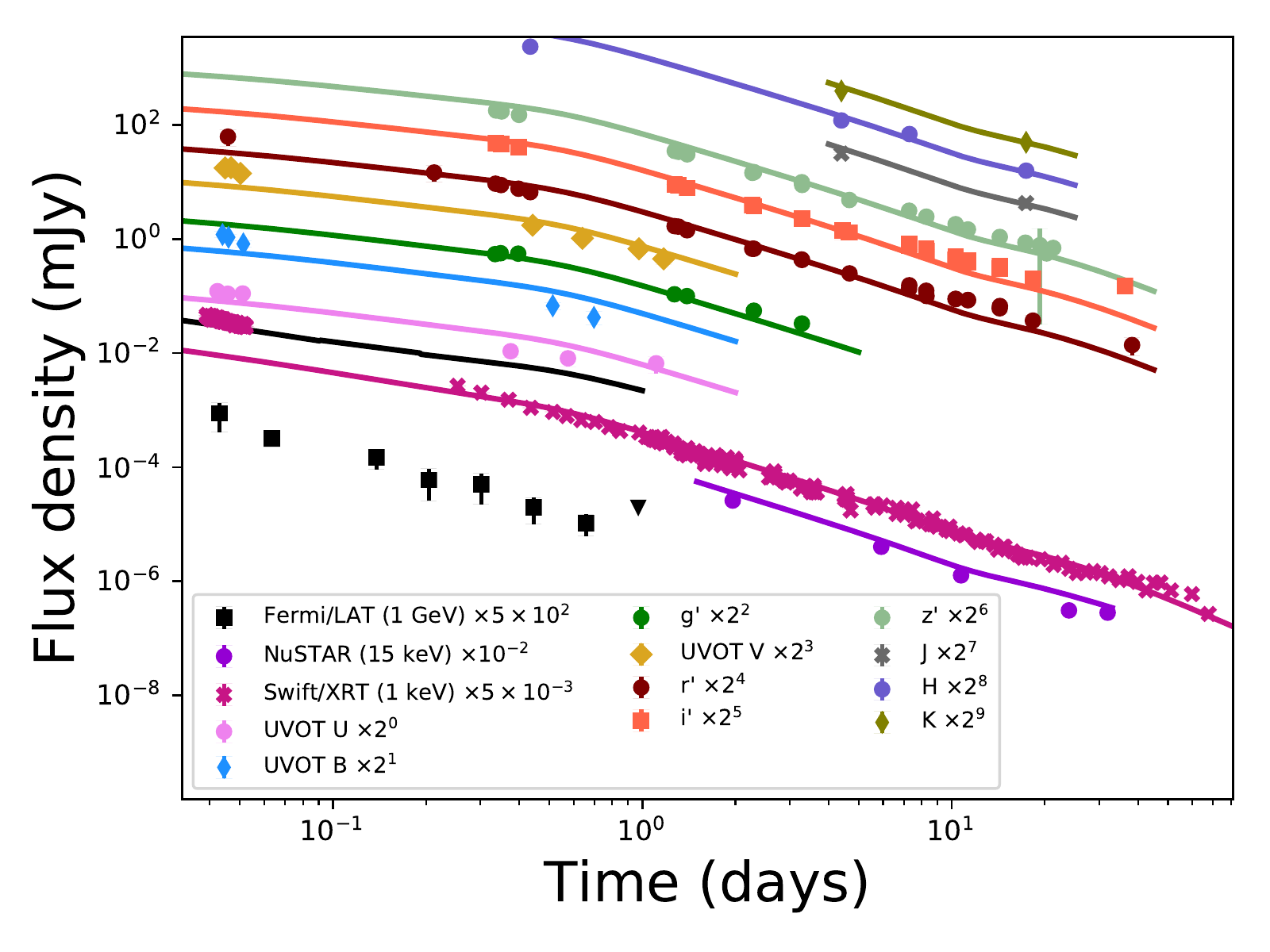}
    \end{tabular}
    \caption{Radio SEDs (left) and light curves (right) for a low-$p$ ISM model with parameters, $p=1.5$, $\epse=0.07$, $\epsb=0.03$, $\dens=10$\,\pcc, $\EKiso=1.5\times10^{52}$~erg, and $A_{\rm V, host}=0.3$~mag. This model requires a jet break at and $\tjet=0.55$~days (implying $\thetajet=9^\circ$ and $\EK=2\times10^{50}$~erg) in order to match the X-ray and optical light curves; however, the model significantly over-predicts the mm-band observations at $\lesssim28$~days, {over-predicts the \fermi/LAT observations by two orders of magnitude,} and does not match the cm-band SED at all, and is, therefore, ruled out.  
    }
    \label{fig:ISM}
\end{figure*}
In this section, we consider an ISM model with $p<2$ in the regime $\nuc<\nuopt<\nux$ with an early jet break ($\tjet\approx0.55$~days) in order to match the optical and X-ray spectral index and light curves, as discussed in Section~\ref{text:fs}. The low value of $p$ is required to match the observed NIR-to-X-ray spectral index of $\betanirx\approx-0.70$ in the stipulated regime of $\nuc<\nuopt<\nux$. We set $p=1.5$ and tune the parameters to match the output X-ray and optical light curves. To calculate the light curves, we replace $\bar{\epsilon}_e$ in \cite{gs02} by $\epse$, and note that there will need to be a change in electron spectrum at some high Lorentz factor in order to keep the total energy in accelerated particles finite. In this model, $\nuc\approx3\times10^{13}~{\rm Hz}<\nuopt$, as required in order to satisfy $\betaniropt\approx\betanirx$ (Section~\ref{text:fs}). We find that in this model the jet becomes non-relativistic at $t_{\rm NR}\approx11$~days and the resultant model light curves over-predict the radio SEDs (Figure~\ref{fig:ISM}). This is fundamentally because it is not possible to match the radio SED onto the optical with a single synchrotron emission component without invoking additional spectral breaks. Thus, the ISM, single-jet model is ruled out for this burst. 

\section{Newtonian RS Model}
\label{appendix:RS}
\begin{figure*}
    \centering    
    \begin{tabular}{cc}
        \includegraphics[width=\columnwidth]{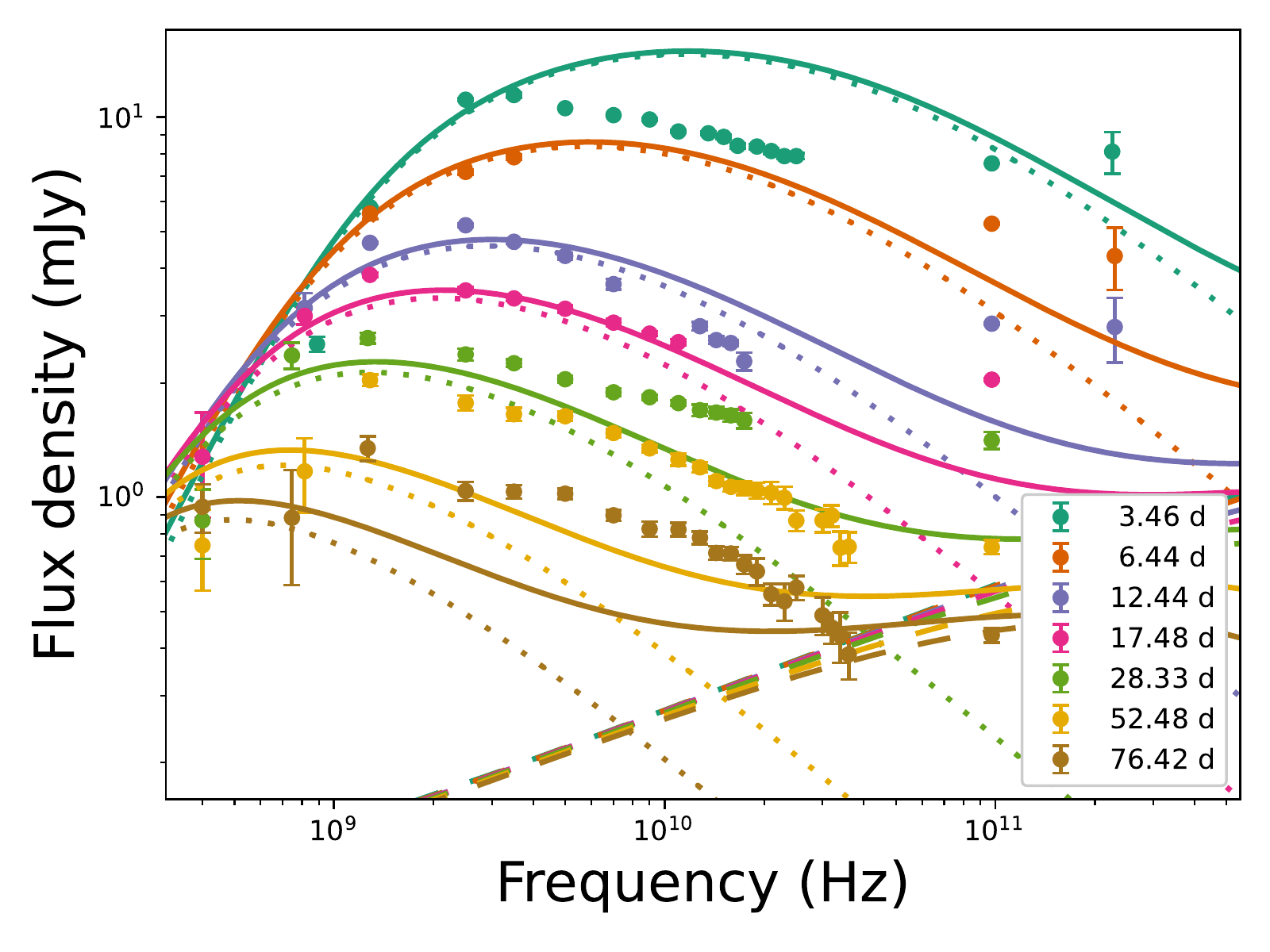} &
        \includegraphics[width=\columnwidth]{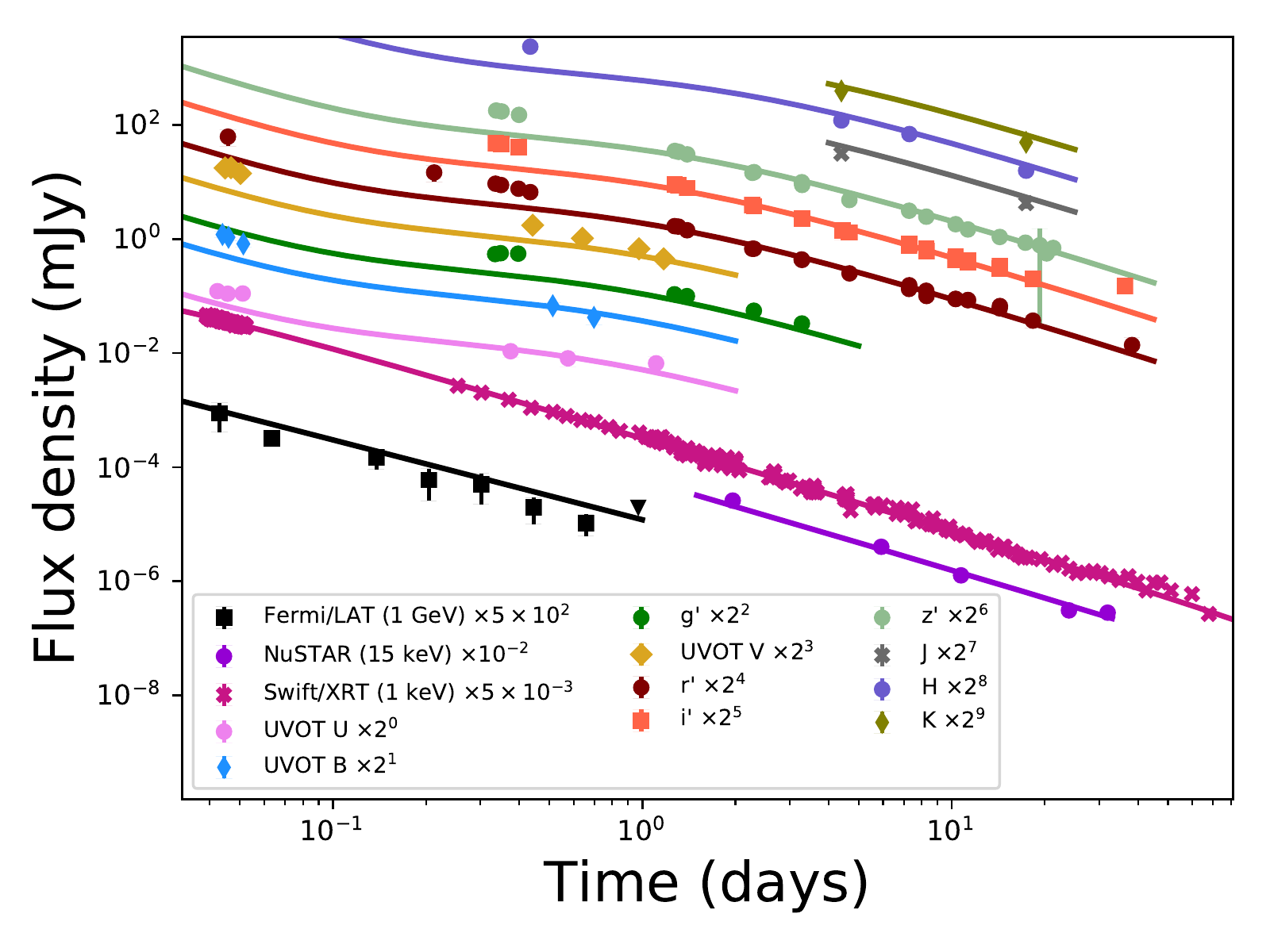}
    \end{tabular}
    \caption{Radio SEDs (left) and light curves (right) for a joint wind model (solid lines) combining a reverse shock (dotted) with the forward shock model (dashed) described in Section~\ref{text:fs}. This Newtonian RS model with $g\approx5$ is tailored to roughly match the cm-band SED at $\approx17.48$~days. Even with an extreme value of $g$, the RS model cannot match the rest of the radio data and {the early ($\lesssim1~d$)} UV/optical light curves, and is, therefore, disfavored. 
    }
    \label{fig:rs}
\end{figure*}
The excess radio emission described in Section~\ref{text:radiocomponents} cannot be easily ascribed to RS emission. To demonstrate this, we combine the FS model described in Section~\ref{text:fs} with an RS model with the following parameters: $\nuar\approx2.3\times10^{9}$~Hz,
$\numr\approx1.2\times10^{11}$~Hz and $\fnumr\approx94$~mJy, selected to achieve an approximate match to the cm-band SED at $\approx17.48$~days. The RS cooling break is only weakly constrained to $\nucr(1\,{\rm day})\gtrsim10^{12}$~Hz in this model, so as to not strongly affect the cm-band SED. We also require $p\approx2$ in order to match the shallow cm-band spectral index above the peak at $\gtrsim2$~GHz. We present a Newtonian RS model with $g=5$ in Figure~\ref{fig:rs}. A higher value of $g$ leads to a slower RS evolution. For a wind medium, we expect $0.5\le g \le 1.5$. Even with  $g\approx5$, the evolution of the model SED {(with a peak given by $\numr$)} is not slow enough to match the observations. A relativistic RS would evolve even faster. Thus, the presently available suite of RS models cannot match the radio observations for this burst. 

\bibliography{GRB221009A_astroph_v2}
\bibliographystyle{aasjournal}

\end{document}